\documentclass[twocolumn,trackchanges]{aastex701}

\usepackage{amsmath,bigdelim}
\usepackage{pbox, cellspace}
\usepackage[dvipsnames]{xcolor}
\usepackage{soul}

\def\kmsec{$\,\rm{km\,s^{-1}}$}

\begin{document}

\title{A 14-year-old Mystery: The Peculiar Case of the Engine-driven SN\,2012ap}

\correspondingauthor{Itai Sfaradi}
\email{itai.sfaradi@berkeley.edu}

\author[0000-0003-0466-3779]{Itai Sfaradi}
\affiliation{Department of Astronomy, University of California, Berkeley, CA 94720-3411, USA}
\affiliation{Berkeley Center for Multi-messenger Research on Astrophysical Transients and Outreach (Multi-RAPTOR), University of California, Berkeley, CA 94720-3411, USA}
\email{itai.sfaradi@berkeley.edu}

\author[0000-0003-4768-7586]{Raffaella Margutti}
\affiliation{Department of Astronomy, University of California, Berkeley, CA 94720-3411, USA}
\affiliation{Berkeley Center for Multi-messenger Research on Astrophysical Transients and Outreach (Multi-RAPTOR), University of California, Berkeley, CA 94720-3411, USA}
\affiliation{Department of Physics, University of California, 366 Physics North MC 7300, Berkeley, CA 94720, USA}
\email{rmargutti@berkeley.edu}

\author[0000-0002-7706-5668]{Ryan Chornock}
\affiliation{Department of Astronomy, University of California, Berkeley, CA 94720-3411, USA}
\affiliation{Berkeley Center for Multi-messenger Research on Astrophysical Transients and Outreach (Multi-RAPTOR), University of California, Berkeley, CA 94720-3411, USA}
\email{chornock@berkeley.edu}

\author[0000-0002-8070-5400]{Nayana A. J.}
\affiliation{Department of Astronomy, University of California, Berkeley, CA 94720-3411, USA}
\affiliation{Berkeley Center for Multi-messenger Research on Astrophysical Transients and Outreach (Multi-RAPTOR), University of California, Berkeley, CA 94720-3411, USA}
\email{nayana@berkeley.edu}

\author[0000-0000-0000-0000]{Eli Wiston}
\affiliation{Department of Astronomy, University of California, Berkeley, CA 94720-3411, USA}
\affiliation{Berkeley Center for Multi-messenger Research on Astrophysical Transients and Outreach (Multi-RAPTOR), University of California, Berkeley, CA 94720-3411, USA}
\email{ewiston@berkeley.edu}

\author[0000-0002-3137-4633]{Fabio De Colle} \affiliation{Instituto de Ciencias Nucleares, Universidad Nacional Aut{\'o}noma de M{\'e}xico, A. P. 70-543 04510 D. F. Mexico}
\email{fabio@nucleares.unam.mx}

\author[0000-0001-6812-7938]{Tracy E. Clarke}
\affiliation{U.S. Naval Research Laboratory, 4555 Overlook Avenue SW, Washington, DC 20375, USA}
\email{tracy.e.clarke2.civ@us.navy.mil}

\author[0000-0002-5187-7107]{Wendy M. Peters}
\affiliation{U.S. Naval Research Laboratory, 4555 Overlook Avenue SW, Washington, DC 20375, USA}
\email{wendy.m.peters8.civ@us.navy.mil}

\author[0000-0001-7833-1043]{Paz Beniamini}
\affiliation{Department of Natural Sciences, The Open University of Israel, PO Box 808, Ra’anana 4353701, Israel} 
\affiliation{Astrophysics Research Center of the Open university (ARCO), The Open University of Israel, PO Box 808, Ra’anana 4353701, Israel} 
\affiliation{Department of Physics, The George Washington University, 725 21st Street NW, Washington, DC 20052, USA}
\email{paz.beniamini@gmail.com}

\author[0000-0002-1568-7461]{Wenbin Lu}
\affiliation{Department of Astronomy, University of California, Berkeley, CA 94720-3411, USA}
\affiliation{Berkeley Center for Multi-messenger Research on Astrophysical Transients and Outreach (Multi-RAPTOR), University of California, Berkeley, CA 94720-3411, USA}
\email{wenbinlu@berkeley.edu}

\author[0000-0002-5565-4824]{Rodolfo Barniol Duran}
\affiliation{Department of Physics and Astronomy, California State University, Sacramento, 6000 J Street, Sacramento CA 95819-6041 USA}
\email{barniolduran@csus.edu}

\author[0000-0002-0592-4152]{Michael Bietenholz}
\affiliation{Department of Physics and Astronomy, York University, Toronto, M3J~1P3, Ontario, Canada}
\affiliation{Hartebeesthoek Radio Astronomy Observatory, PO Box 443, Krugersdorp, 1740, South Africa}
\email{mbieten@yorku.ca}

\author[0000-0003-0528-202X]{Collin~T.~Christy}
\affiliation{Department of Astronomy and Steward Observatory, University of Arizona, 933 North Cherry Avenue, Tucson, AZ 85721-0065, USA}
\email{collinchristy@arizona.edu}

\author[0000-0001-5126-6237]{Deanne L. Coppejans} \affiliation{Department of Physics, University of Warwick, Gibbet Hill Road, Coventry, CV4 7AL, UK}
\email{deanne.coppejans@warwick.ac.uk}

\author[0000-0001-7081-0082]{Maria R. Drout}
\affiliation{David A. Dunlap Department of Astronomy \& Astrophysics, University of Toronto, 50 St. George St., Toronto, ON M5S 3H4, Canada}
\email{maria.drout@utoronto.ca}

\author[0009-0008-0782-5028]{Dina Ibrahimzade}\affiliation{Department of Astronomy, University of California, Berkeley, CA 94720-3411, USA}
\email{Dinaevazade@gmail.com}

\author[0000-0001-9033-4140]{Micha{\l} J.~Micha{\l}owski}
\affiliation{Astronomical Observatory Institute, Faculty of Physics and Astronomy, Adam Mickiewicz University, ul. S{\l}oneczna 36, 60-286, Pozna\'{n}, Poland}
\email{michal.michalowski@amu.edu.pl}

\author[0000-0002-0763-3885]{Dan Milisavljevic} \affiliation{Department of Physics and Astronomy, Purdue University, 525 Northwestern Avenue, West Lafayette, IN 47907, USA}
\email{dmilisav@purdue.edu}

\author[0000-0002-9646-8710]{Conor M. B. Omand}
\affiliation{Astrophysics Research Institute, Liverpool John Moores University, Liverpool Science Park IC2, 146 Brownlow Hill, Liverpool, UK, L3 5R}
\email{C.M.Omand@ljmu.ac.uk}

\author[0000-0002-8614-8721]{Yihan Wang}
\affiliation{Department of Astronomy, University of Wisconsin, Madison, WI 53706, USA.}
\email{wang3697@wisc.edu}

\author[0000-0002-8297-2473]{Kate D.~Alexander}
\affiliation{Department of Astronomy and Steward Observatory, University of Arizona, 933 North Cherry Avenue, Tucson, AZ 85721-0065, USA}
\email{kdalexander@arizona.edu}

\author[0000-0003-3494-343X]{Carles Badenes}
\affiliation{Department of Physics and Astronomy and Pittsburgh Particle Physics, Astrophysics and Cosmology Center (PITT PACC), University of Pittsburgh, 3941 O'Hara Street, Pittsburgh, PA 15260, USA}
\email{badenes@pitt.edu}

\author[0000-0002-7735-5796]{Joe Bright}
\affiliation{Astrophysics, Department of Physics, University of Oxford, Keble Road, Oxford, OX1 3RH, UK}
\email{joe.bright@physics.ox.ac.uk}

\author[0000-0001-8530-8941]{Jonathan Granot}
\affiliation{Department of Natural Sciences, The Open University of Israel, PO Box 808, Ra’anana 4353701, Israel} \affiliation{Astrophysics Research Center of the Open university (ARCO), The Open University of Israel, PO Box 808, Ra’anana 4353701, Israel} \affiliation{Department of Physics, The George Washington University, 725 21st Street NW, Washington, DC 20052, USA}
\email{granot@openu.ac.il}

\author[0000-0002-5698-8703]{Erica Hammerstein}
\affiliation{Department of Astronomy, University of California, Berkeley, CA 94720-3411, USA}
\affiliation{Berkeley Center for Multi-messenger Research on Astrophysical Transients and Outreach (Multi-RAPTOR), University of California, Berkeley, CA 94720-3411, USA}
\email{ekhammer@berkeley.edu}

\author[0000-0002-3934-2644]{W.~V.~Jacobson-Gal\'{a}n}
\altaffiliation{NASA Hubble Fellow}
\affiliation{Cahill Center for Astrophysics, California Institute of Technology, MC 249-17, 1216 E California Boulevard, Pasadena, CA, 91125, USA}
\email{wynnjg@caltech.edu}

\author[0000-0002-2249-0595]{Natalie LeBaron}
\affiliation{Department of Astronomy, University of California, Berkeley, CA 94720-3411, USA}
\affiliation{Berkeley Center for Multi-messenger Research on Astrophysical Transients and Outreach (Multi-RAPTOR), University of California, Berkeley, CA 94720-3411, USA}
\email{nlebaron@berkeley.edu}

\author[0000-0002-5358-5642]{Kohta Murase}
\affiliation{Center for Gravitational Physics and Quantum Information, Yukawa Institute for Theoretical Physics, Kyoto, Kyoto 606-8502 Japan}
\affiliation{Department of Physics, Department of Astronomy and Astrophysics, Center for Multimessenger Astrophysics, Institute for Gravitation and the Cosmos, The Pennsylvania State University, University Park, PA 16802, USA}
\email{murase@psu.edu}

\author[0009-0009-4872-1134]{Gitika Rameshan} 
\affiliation{Indian Institute of Astrophysics, II Block, Koramangala, Bengaluru-560034, Karnataka, India} 
\affiliation{Academy of Scientific and Innovative Research (AcSIR), Ghaziabad, Uttar Pradesh, 201002, India}
\email{gitikaramesan@gmail.com}

\author[0000-0001-8023-4912]{Huei Sears}
\affiliation{Department of Physics and Astronomy, Rutgers, The State University of New Jersey, 136 Frelinghuysen Road, Piscataway, NJ 08854-8019, USA}
\email{huei.sears@rutgers.edu}

\author[0000-0002-3019-4577]{Michael Stroh}
\affiliation{National Radio Astronomy Observatory; 5651 Balloon Fiesta Pkwy. NE; Albuquerque, NM 87113}
\email{michael.c.stroh@gmail.com}

\author[0000-0003-0794-5982]{Giacomo Terreran}
\affiliation{Adler Planetarium, 1300 S DuSable Lake Shore Dr, Chicago, IL 60605, USA}
\email{gqterre@gmail.com}

\begin{abstract}
We present late-time ($\delta t > 3000$\,d) optical (Keck), X-ray (Chandra and NuSTAR), and radio (VLA, ALMA, and the uGMRT) observations of the Type Ic-BL SN\,2012ap. Previous studies of this SN suggested that it stands out as a key example of a weak engine-driven explosion due to the lack of gamma-ray burst detection and a mildly relativistic ejecta. Recently, radio sky surveys revealed the rebrightening of the radio emission from this SN, highlighting the possibilities of a density enhancement at large radii or the existence of an off-axis relativistic jet. While the late-time optical spectra does not exhibit the broad emission lines seen in other interacting SNe, our analysis of the broadband radio and X-ray emission implies that both scenarios are plausible. If a density enhancement is responsible for the radio rebrightening, it has to result from a change in the mass-loss rate and/or wind velocity, possibly due to the transition of the progenitor from a red supergiant to a Wolf-Rayet star. If the late-time radio component is a result of an off-axis relativistic jet, we find that an energetic narrow jet viewed at $\theta_{\rm obs} \geq 80^{\circ}$ is needed. In this scenario, SN\,2012ap is not a result of a weak engine-driven explosion, and, instead, it is similar to other GRBs. However, radio rebrightenings of Type Ic-BL SNe are not enough on their own to determine the existence of off-axis jets and our planned VLBA observation will help reveal the true nature of this SN.
\end{abstract}


\section{Introduction} 
\label{sec: intro}

Core-collapse supernova (CCSN) explosions are the final fate of massive stars and typically result in high-velocity ejecta (\citealt{Smartt_2009, Jerkstrand_2026} and the references therein). Relativistic SNe are a subset of these explosions in which the SN shock front reaches mildly relativistic velocities (with forward shock velocities, $v_{\rm sh} \geq 0.5 c$). A central engine is often introduced as the mechanism that accelerates the ejecta to these velocities \citep{MacFadyen_2001, Berger_2003, Corsi_2021}. Observationally, some relativistic SNe are associated with long-duration gamma-ray bursts (GRBs; \citealt{Galama_1998, Stanek_2003, Piran_2004, Soderberg_2006, Woosley_2006}), possibly powered by shocks from relativistic jets propagating through the progenitor Wolf-Rayet star \citep{Kumar_2015}. The propagation of the jet through the massive progenitor star can form a wide angle structure, resulting in a jet+cocoon system \citep{Lazzati_2005, Nakar_2017, Corsi_2021, De_Colle_2022}. Therefore, these explosions provide key observational constraints on the physical conditions that determine the jet formation and evolution.

Radio emission from SNe is typically associated with synchrotron radiation generated in shock waves formed by the interaction of fast-moving ejecta with the ambient medium \citep{Sari_1998, Chevalier_1998}. Therefore, observing at radio wavelengths is sometimes the only probe of the shock evolution, its environment, and the mass-loss processes that formed this environment. Analysis of large samples of radio observations of CCSNe (e.g., \citealt{Weiler_2002, Soderberg_2006, Bietenholz_2021, Stroh_2021, Rose_2024, Sfaradi_2025}) revealed a wide-variety of rise and fall times of the radio emission, radio luminosities, shock velocities, density profiles, and mass-loss rates from their progenitors. More specifically, the observed radio emission from SNe of type Ic-BL (some associated with long-duration GRBs) reveals a fast rise time to peak, $t_{\rm pk} \sim 10$\,d after the explosion, radio luminosities as high as $L_{\rm pk} \sim 10^{27} \, \rm erg \, s^{-1} \, Hz^{-1}$, and, sometimes, mildly relativistic shock velocities \citep{Berger_2003, Soderberg_2006, Soderberg_2010, Chakraborti_2015}.

Some CCSNe exhibit late-time radio (re-)brightening on timescales of months to years after optical discovery. These are usually interpreted as either the interaction of the SN ejecta with a complex CSM structure, an off-axis jet entering our line of sight, or pulsar wind nebula. For example, the interaction of the ejecta of SN\,2014C, with the complex CSM structure around its progenitor \citep{Milisavljevic_2015_14c, Margutti_2017_14c, Brethauer22} resulted in a double-peaked radio light curve \citep{Anderson_2017}. Very Long Baseline Interferometry (VLBI) observations of this source confirmed the non-relativistic nature of the radio emitting source \citep{Bietenholz_2018}. Type Ic-BL SNe are particularly interesting in this context as they are sometimes associated with long-duration GRBs and, therefore, make perfect candidates for off-axis jets. For example, the double-peaked radio light curve of PTF11qcj was interpreted as either due to late-time CSM interaction or an off-axis jet \citep{Corsi_2014, Palliyaguru_2019}. Then, direct measurements of the size of the source with VLBI observations revealed an average shock velocity of $\sim 0.036c$ \citep{Palliyaguru_2021}, ruling out the existence of a relativistic jet. Recently, \cite{Scheroder_2025} suggested that the late-time radio emission from the Type Ic-BL SN PTF10tqv is consistent with an off-axis jet entering our line of sight; however, the radio light curve alone cannot be used to rule out CSM interaction at large radii.

\subsection{The relativistic SN\,2012ap}
\label{subsec: SN2012ap_intro}

Among relativistic SNe, SNe\,2009bb \citep{Soderberg_2010, Pignata_2011} and 2012ap \citep{Margutti_2014, Milisavljevic_2015, Chakraborti_2015,Zheng_2015} stand out as the only two relativistic SNe without a GRB detection. These events have been interpreted as the weakest members of the engine-driven explosion family, where a central engine launches a jet that fails to fully break out of the progenitor envelope, implying a continuum of explosions between the spherical non-relativistic SNe and the fully collimated relativistic GRBs.

In this paper we focus on the late-time radio, optical, and X-ray emission from SN\,2012ap, which was first discovered on 2012 February 10, by the Lick Supernova Search \citep{Jewett_2012}. SN\,2012ap is located in the galaxy NGC 1729, and we adopt here the luminosity distance to the galaxy, $d_L=40$ Mpc \citep{Margutti_2014}, as the distance of the SN. Previous modeling of the optical emission from this SN suggested that the explosion date is 2012 February 5 (MJD 55962; \citealt{Milisavljevic_2015}), and we define here $\delta t$ as the time (in the frame of the observer) since this explosion date. 

Early radio observations of SN\,2012ap (from $\delta t = 12$\,d to $38$\,d) revealed a bright source with $L_{\rm \nu} \sim 10^{28} \, \rm erg \, s^{-1} \, Hz^{-1}$ and a broken-power-law spectral energy distribution (SED). \cite{Chakraborti_2015} associated this emission with a mildly relativistic outflow traveling in a wind-like CSM. Recent radio observations of SN\,2012ap \citep{Stroh_2021, Rose_2024} revealed rebrightening of the GHz emission at very late-times (up to $10$ years after the SN explosion), suggestive of either (i) a density enhancement at large radii around the SN, or (ii) relativistic jet observed off-axis, or (iii) a newly formed pulsar wind nebula (PWN). We present here new late-time, panchromatic observations of the relativistic SN\,2012ap that are essential to understand the nature of this SN (X-ray in \S~\ref{sec:XrayObs}, optical in \S~\ref{sec: optical_observations}, and broadband radio in \S~\ref{sec: radio_observations}). We analyze our observations in \S~\ref{sec: sn2012ap_radio} and discuss our results in \S~\ref{sec: disc}. \S~\ref{sec: conc} is for conclusions.

\section{X-ray observations} 
\label{sec:XrayObs}

\subsection{Chandra (0.3--10 keV)}
\label{SubSec:Chandra}
SN\,2012ap was observed with the Chandra X-ray Observatory (CXO) at $\delta t\approx24\,$d and no source was detected down to a luminosity of $L_{x} < 2.4 \times 10^{39} \, {\rm erg \, s^{-1}}$ \citep[in the range of 0.3--10\,keV]{Margutti_2014}. We obtained three additional epochs of deep X-ray CXO observations at $\delta t=3596$\,d, 3999\,d, and 4082\,d (observation IDs 25181, 25216, 25217; PIs Stroh and Chornock). We analyzed the ACIS-S data following standard practice with \texttt{CIAO v4.15} and corresponding calibration files. A source of X-ray emission is blindly detected with \texttt{wavdetect} at the optical location of SN\,2012ap in all epochs with significance and net count-rates reported in Table \ref{Tab:Xraydata}. For each ID we used \texttt{specextract} to extract a spectrum using a source region with radius of $1.5\arcsec$ and a source-free background region. We fitted the spectra with an absorbed power-law model. The neutral hydrogen absorption column in the direction of the transient is $\rm{NH_{MW}}=4.9\times 10^{20}\,\rm{cm^{-2}}$ \citep{HI4PI}. Our spectral analysis (see below) revealed no evidence for intrinsic absorption, and we therefore assumed $\rm{NH}_{\rm{int}}=0\,\rm{cm^{-2}}$ in our analysis.  The best-fitting parameters and inferred fluxes for a power-law spectral  model are reported Table \ref{Tab:Xraydata}. Because of the limited count statistics we cannot differentiate between spectral models. Relevant to the discussion above, we note that a joint fit of all late-time Chandra data with a thermal bremsstrahlung model leads to a temperature constraint of $10\lesssim T\lesssim 20$\,keV and normalization $\rm{Norm}=2.6^{+0.6}_{-0.4}\times 10^{-6}$. In \texttt{Xspec} the normalization of the bremsstrahlung model is defined as $\rm{Norm}\equiv\frac{3.02\times 10^{-15}}{4\pi d_L^2  }\int  n_e n_I dV$, where $n_e$ and $n_I$  are the electron and ion number densities of the emitting region (in the units of $\rm{cm^{-3}}$), respectively, and $d_L/\rm{cm}$ is the distance to the target. 

In general, irrespective of the spectral model, we find evidence for a brightening of the source at $\delta t \gtrsim 9.8$\,yr compared to the initial month post explosion, with an X-ray luminosity $L_x\approx 3\times 10^{39}\,\rm{erg\,s^{-1}}$ that is consistent with being constant in the time period  $\delta t = 9.8-11.2$\,yr.  

\subsection{NuSTAR (3--80 keV)}
\label{SubSec:NuSTAR}

We obtained two epochs of hard X-ray observations with the Nuclear Spectroscopic Telescope (NuSTAR) in coordination with CXO observations at $\delta t = 3998$ and $4081$\,d (Obs IDs 80802504002 and 80802504004, PI Chornock). We analyzed the data with the NuSTAR Data Analysis Software (v1.9.7) and calibration files. A first extraction with  \texttt{nupipeline} and standard filtering led to non-detections. To obtain the best constraints on the source flux, we removed the time intervals with high background largely due to solar flares using custom scripts. We report the count-rate limits in the 8--20 keV energy band (which is around the peak of the NuSTAR effective area) combining the signal from both modules in Table \ref{Tab:Xraydata}. We assume a photon index $\Gamma=2$ for the flux calibration as indicated by Chandra observations.

\section{Optical spectroscopy} 
\label{sec: optical_observations}

\begin{figure*}
\centering
    \includegraphics[width=0.6\textwidth]{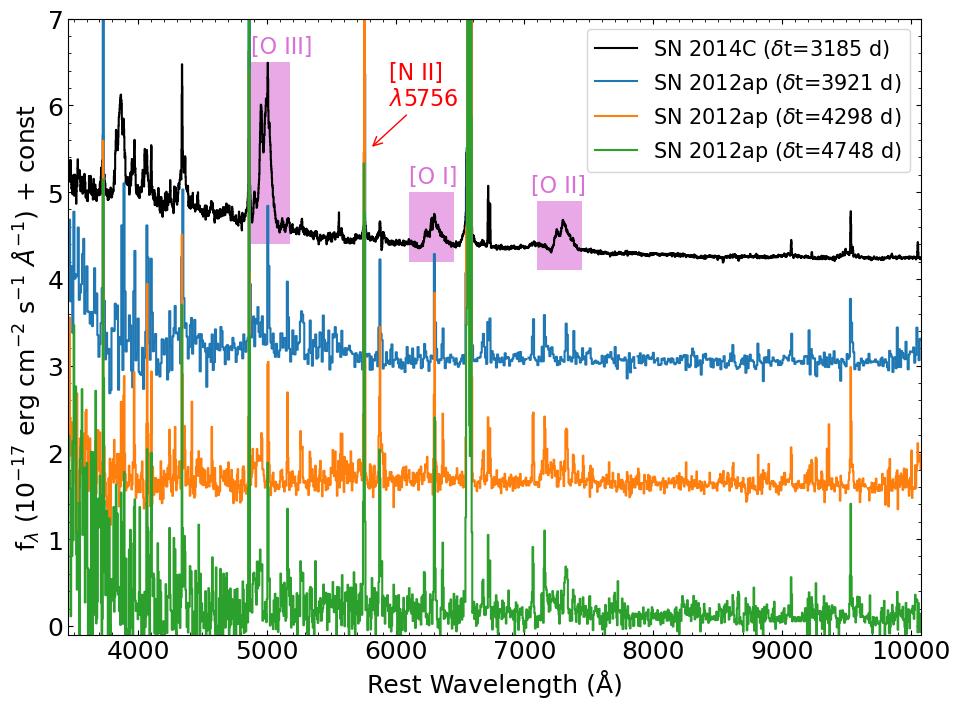} 
    \includegraphics[width=0.35\textwidth]{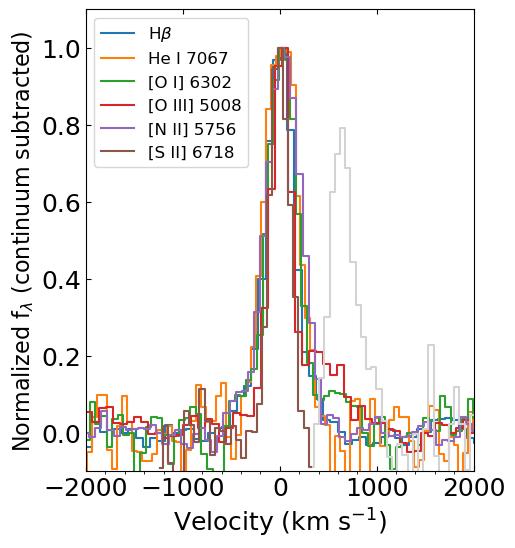}
    \caption{{\it Left:} LRIS spectral sequence of SN\,2012ap compared to the interacting SN\,2014C at late times. The purple shaded regions mark the broad (FWHM $\sim$5000 \kmsec) oxygen emission lines from ejecta illuminated by the CSM interaction region in SN\,2014C. These are not present in SN\,2012ap, while narrow [\ion{N}{2}] $\lambda$5756 emission is unusually prominent. All spectra have been dereddened by assumed values of $E(B-V)$=0.75~mag (SN\,2014C; \citealt{milisav14c}) and 0.6~mag (SN\,2012ap; \citealt{Milisavljevic_2015}). {\it Right:} Line profiles in the $\delta$t=4298~d spectrum of SN\,2012ap, demonstrating a lack of broad (FWHM$\gtrsim$1000 \kmsec) emission from SN ejecta. The [\ion{S}{2}] $\lambda$6733 line is grayed out for clarity.}
    \label{fig:optspec}
\end{figure*}

We obtained three epochs of late-time optical spectroscopy of SN\,2012ap on 2022 Oct. 31 ($\delta t =3921$\,d), 2023 Nov. 11 ($\delta t = 4298$\,d), and 2025 Feb. 04 ($\delta t = 4748$\,d) using the Low-Resolution Imaging Spectrometer (LRIS; \citealt{lrisref}) on the Keck-I 10~m telescope to cover the full optical spectral range from 3100--10300~\AA. All observations used the 1$\arcsec$ slit, the D560 dichroic beamsplitter, and the 400/8500 grism on the red side, resulting in a spectral resolution of $\sim$6.5~\AA. The 2022 spectrum was taken with the 400/3400 grism on the blue side, with a resolution of $\sim$6~\AA, while the other two epochs used the 600/4000 grism, with resolutions of $\sim$4~\AA. A source is well detected at the position of SN\,2012ap in all observations, which is sufficiently offset from its host galaxy that background subtraction was straightforward. Spectral reduction and extraction followed standard procedures, as outlined by \citet{silverman2012}. All wavelengths are quoted in vacuum. The spectral sequence is presented in Figure~\ref{fig:optspec}.

In addition, we present a previously unpublished late-time LRIS optical spectrum  of SN\,2014C, a prototypical example of a supernova exhibiting delayed CSM interaction \citep{milisav14c,margutti14c}. This spectrum was acquired on 2022 Sep. 25 and reduced in a similar manner to the SN 2012ap data. SN\,2014C exhibited photospheric spectra typical of SNe Ib, but by $\sim$100~d after explosion, interaction with H-rich CSM was evident from intermediate-width (FWHM$\approx$1200 \kmsec) H$\alpha$ emission produced in shocked CSM \citep{milisav14c}. As the object evolved, the H$\alpha$ emission became less prominent and strong, broad (FWHM$\approx$5000 \kmsec) forbidden emission lines from the SN ejecta became more prominent ([\ion{O}{1}], [\ion{O}{2}], and [\ion{O}{3}] are marked on Fig.~\ref{fig:optspec}; \citealt{mauerhan18,thomas22,tinyanont25}). As the input from radioactive decay has faded on these decade timescales, the SN ejecta emission must be powered by another source, believed to be reprocessing of X-rays from the reverse shock.

By contrast, no broad or intermediate-width emission lines, either from the SN ejecta or hydrogen in the CSM, are securely detected for SN\,2012ap. Instead, a large number of narrow emission lines from a range of ionization states are present. 
The right panel of Figure~\ref{fig:optspec} shows selected line profiles in velocity space. While some low-ionization lines (e.g., [\ion{S}{2}]) have profiles consistent with the instrumental resolution, some of the lines associated with the SN have somewhat broader profiles. For example, the [\ion{N}{2}] $\lambda$5756 emission line has a FWHM of 8.2$\pm$0.1~\AA, compared to the instrumental resolution of $\sim$6~\AA, which requires an additional source of broadening equivalent to $\sim$300 \kmsec\ FWHM. This is still much lower than the velocity widths expected if the emitting material has been shocked by the SN.

The ratios of the narrow lines indicate that the emission is not dominated by an unrelated \ion{H}{2} region and that the emitting nebula is associated with CSM gas near the SN. Most notably, auroral [\ion{N}{2}] $\lambda$5756 emission is stronger than [\ion{O}{2}] $\lambda$3728 or [\ion{O}{3}] $\lambda$5008. This would be unexpected from an \ion{H}{2} region with solar abundance ratios whose central ionizing source is a massive star. However, this temperature-sensitive line also became strong at late times in other SNe experiencing CSM interaction, such as SNe\,2005ip and 2004dk \citep{smith05ip,mauerhan18}.  In addition, the unusually strong [\ion{N}{2}] emission motivates consideration of enhanced nitrogen abundances in future work, potentially from CNO-processed material, as has been previously observed in the CSM of interacting SNe (e.g., \citealt{fransson05}).

As an example of the complexity of the spectrum, the [\ion{S}{2}] line flux ratio of $j_{\lambda6718}$/$j_{\lambda6733}$ is a classic density tracer for \ion{H}{2} regions. The observed value of $1.25\pm0.07$ is consistent with an electron density of $n_e \approx 200 (T/10^4~{\rm K})^{1/2}$~cm$^{-3}$, typical of \ion{H}{2} regions (e.g., \citealt{agn2}). However, at this low density, the [\ion{N}{2}] $\lambda$5756 and [\ion{O}{3}] $\lambda4364$ lines would be much weaker than observed for any value of the temperature. The density of the emitting gas needs to be sufficiently high that collisional deexcitation of the $^1D_2$ levels is not negligible for these ions. The observed (but de-reddened) values of the [\ion{O}{3}] line ratio ($j_{\lambda4960} + j_{\lambda5008}$)/$j_{\lambda 4364}\approx 4.3$ and the [\ion{N}{2}] line ratio ($j_{\lambda6550} + j_{\lambda6585}$)/$j_{\lambda 5756}\approx 1.8$ can only be accommodated if the emitting gas has both a high temperature ($T\gtrsim15,000$~K) and a high electron density, $n_e \approx (0.4-1)\times 10^6$~cm$^{-3}$ \citep{agn2}. On the other hand, we do not detect coronal lines such as [\ion{Fe}{7}] $\lambda$6089 or [\ion{Fe}{10}] $\lambda$6376 that are diagnostic of $10^6$~K gas and are sometimes seen in interacting SNe (e.g., \citealt{turatto93,smith05ip}). 

We defer further photoionization analysis of the optical spectra to future work, particularly as the observed spectra clearly represent the unresolved superposition of multiple zones with different conditions. However, the optical emission from the SN appears indicative of the presence of dense, unshocked, and possibly CNO-enhanced, CSM at large radii from the explosion site.

\section{Radio observations} 
\label{sec: radio_observations}

Broadband radio observations of SN\,2012ap during the first $38$\,d after stellar explosion were first reported and analyzed by \cite{Chakraborti_2015}. Late-time observations (at $\delta t > 2000$\,d) revealing the radio rebrightening of SN\,2012ap were obtained with the Very Large Array All Sky Survey (VLASS; \citealt{VLASS_paper}) and the Australian Square Kilometre Array Pathfinder (ASKAP; \citealt{ASKAP_paper}) Variable And Slow Transient survey (VAST; \citealt{VAST_paper}) \citep{Stroh_2021, Rose_2024}. We obtained broadband radio observations at late-times ($\delta t > 3200$\,d) using the Karl G. Jansky Very Large Array (VLA) and the upgraded Giant Metrewave Radio Telescope (uGMRT). In addition, we report an observation made by the Atacama Large Millimeter/Submillimeter Array (ALMA).

\subsection{The Very Large Array (VLA)}
\label{subsec: VLA}

Following the rebrightening observed in VLASS observations \citep{Stroh_2021} we obtained two broadband ($\sim 1.5-10 \, {\rm GHz}$) observations with the VLA on 2021 January 30 (VLA 20B-279; PI: Stroh), and 2024 October 30 (VLA 24B-311; PI: Wiston). During the first observation, the absolute flux density and bandpass calibrator was 3C138 and ICRF J050112.8-015914 (J0501-0159 hereafter) was used to calibrate the phases. For the second observation, 3C48 was used as an absolute flux and bandpass calibrator, and J0501-0159 as the phase calibrator. We used the Common Astronomy Software Applications (CASA; \citealt{CASA}) packages and the VLA calibration pipeline (v6.5.4.9) to flag and calibrate the data. 

To complement these observations, we used data taken with the VLA Low Band Ionospheric and Transient Experiment (VLITE; \citealt{VLITE_paper}) in lower bands ($\sim 0.35 \, {\rm GHz}$) at the same time as our broadband observations. The VLITE data were calibrated using the standard VLITE calibration pipeline \citep{Polisensky2016}. Baselines shorter than 3 and 6 klambda were removed from the 2021 and 2024 datasets respectively due to radio frequency interference. The data were then self-calibrated in phase and imaged with a robust weighting factor of 0 using the Obit task MFImage \citep{Cotton2008}.

We used the CASA tasks \texttt{TCLEAN} to produce clean images of the field, \texttt{IMSTAT} to calculate the image rms, and \texttt{IMFIT} to fit the point source at the position of the supernova. We then estimate the error of the flux density to be a quadratic sum of the error produced by the CASA task \texttt{IMFIT} and a $10\%$ calibration error \citep{Weiler_1986}. We report the flux density measurements in Table~\ref{tab: radio_observations} in the appendix.

\subsection{The Upgraded Giant Metrewave Radio Telescope}
\label{subsec: uGMRT}

We observed SN\,2012ap with the upgraded Giant Metrewave Radio Telescope (uGMRT) on 2022 August 14.10, 16.12, and 17.10 in the frequency bands 250$-$500 MHz (band-3), 550$-$850 MHz (band-4), and 1050$-$1450 MHz (band-5) respectively (under project code 42\_096 (PI: Nayana A. J.). The data were recorded in full polarization mode with an integration interval of 10 seconds. We use 200 MHz bandwidth centered around $0.33$ and $0.65$ GHz (band-3 and 4, respectively), and 400 MHz centered around $1.37$ GHz (band-5) split into 2048 channels. 3C286 was used as the flux density calibrator. We used source J0447-220 to calibrate the phase corrections. We used Astronomical Image Processing Software (AIPS) to analyze the data. Standard flagging and calibration procedures were followed using AIPS tasks. The calibrated target source data were imaged using the \texttt{IMAGR} task, and we used AIPS task \texttt{JMFIT} to fit the point source. We then estimate the flux density error to be a quadratic sum of the error produced by the AIPS task \texttt{JMFIT} and a calibration error of $10\%$. We report the flux density measurements in Table~\ref{tab: radio_observations} in the appendix.

\subsection{Atacama Large Millimeter/Submillimeter Array}
\label{subsec: ALMA}

ALMA observed the field of SN\,2012ap on 2022 January 1, as part of the ALMA Carbon Monoxide Supernova (ACOS) survey: testing the single-star and binary models of type Ic supernovae (ALMA program ID 2021.1.00099.S; PI: Michałowski; \citealt{Martin_2024}). We used the standard National Radio Astronomy Observatory (NRAO) calibrated images, fitted the point source with CASA task \texttt{IMFIT}, and used CASA task \texttt{IMSTAT} to obtain the image rms. We provide the flux density measurements in Table~\ref{tab: radio_observations}.

\section{The origin of the late-time broadband emission from SN\,2012ap} 
\label{sec: sn2012ap_radio}

\begin{figure}
    \includegraphics[width=\linewidth]{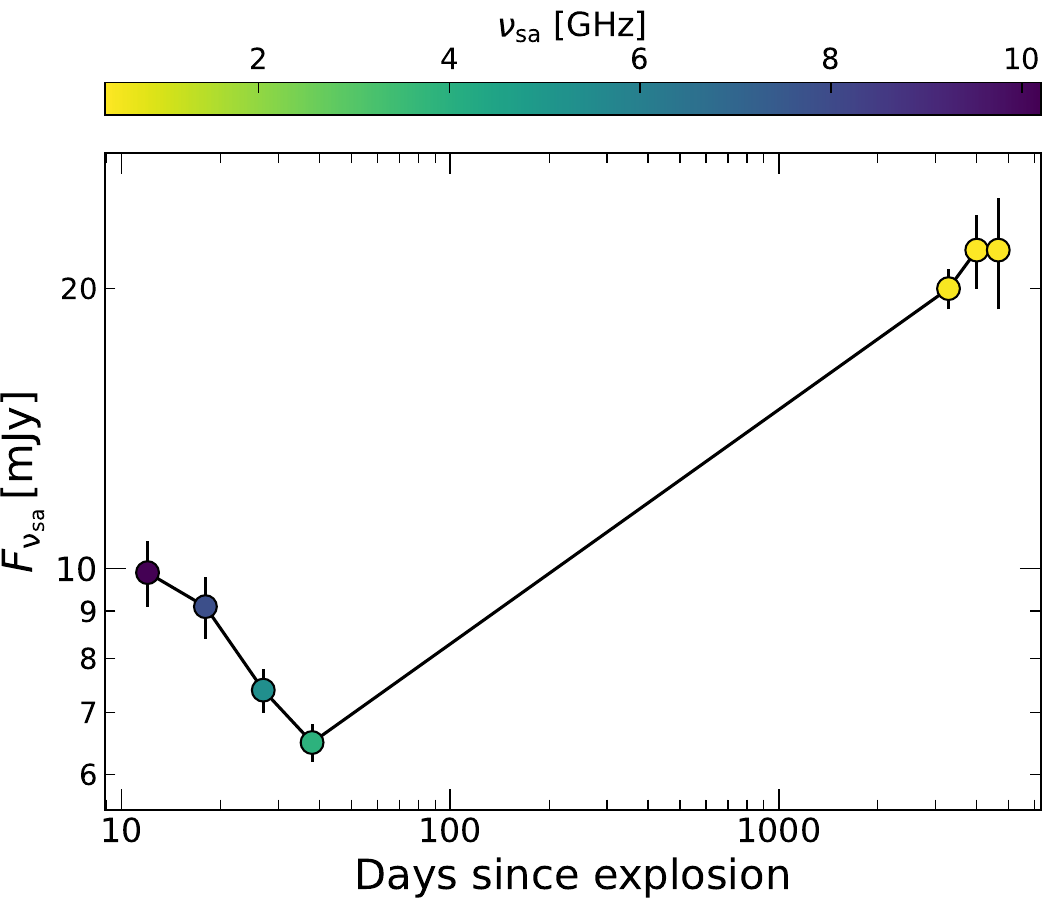}
    \caption{The evolution of the flux density at the radio spectral peak with time. We color-code this plot by the position of peak frequency. These values are inferred from fitting Eq.~\ref{eq: bpl_function} to the broadband radio SEDs (see \S~\ref{subsub:radioFS} for detailed discussion and Table~\ref{tab: bpl_fits_parameters} for the fitted parameters).}
    \label{fig: SN2012ap_peaks}
\end{figure}

\begin{figure*}
    \includegraphics[width=\linewidth]{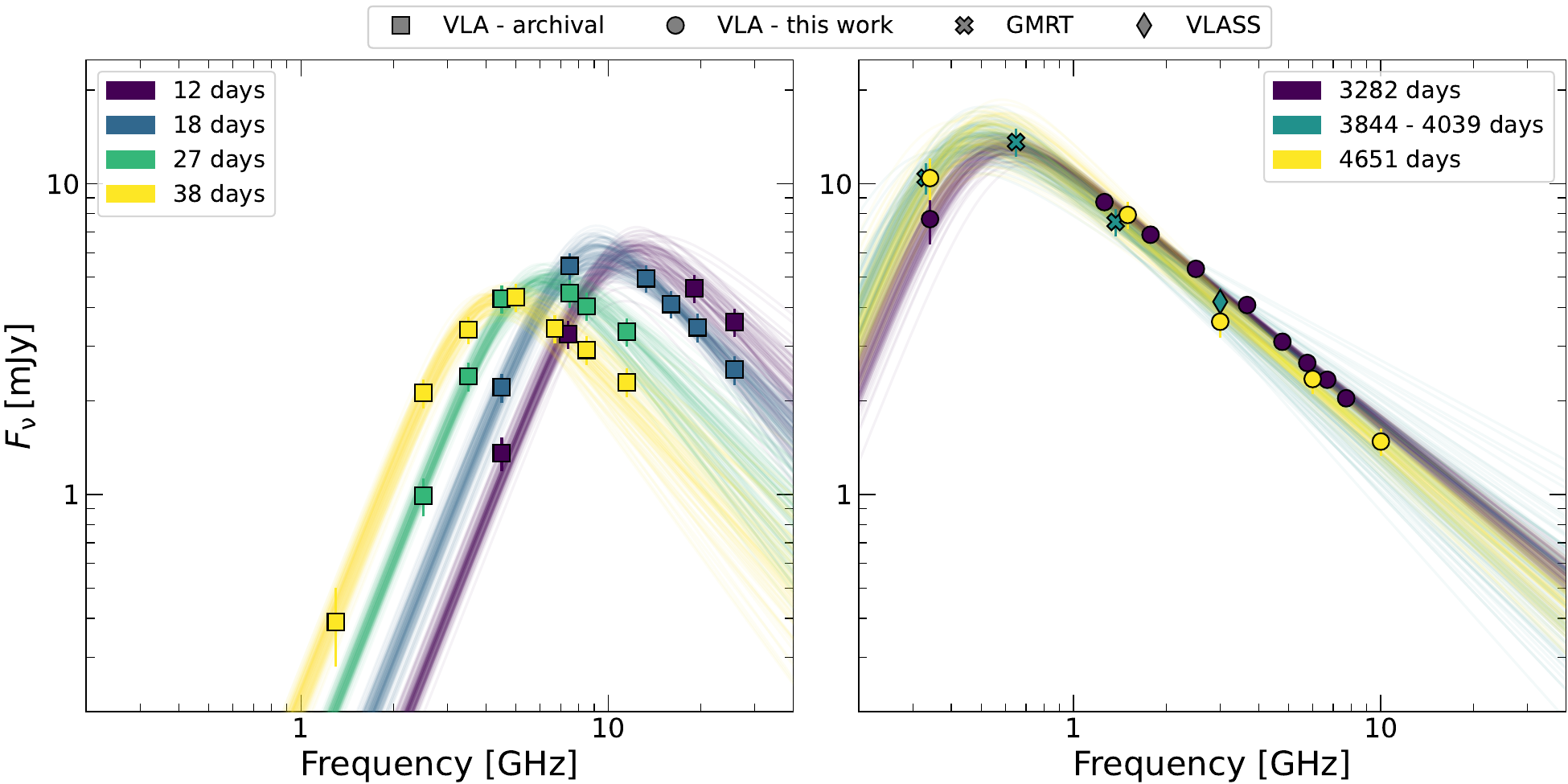} 
    \\
    \includegraphics[width=\linewidth]{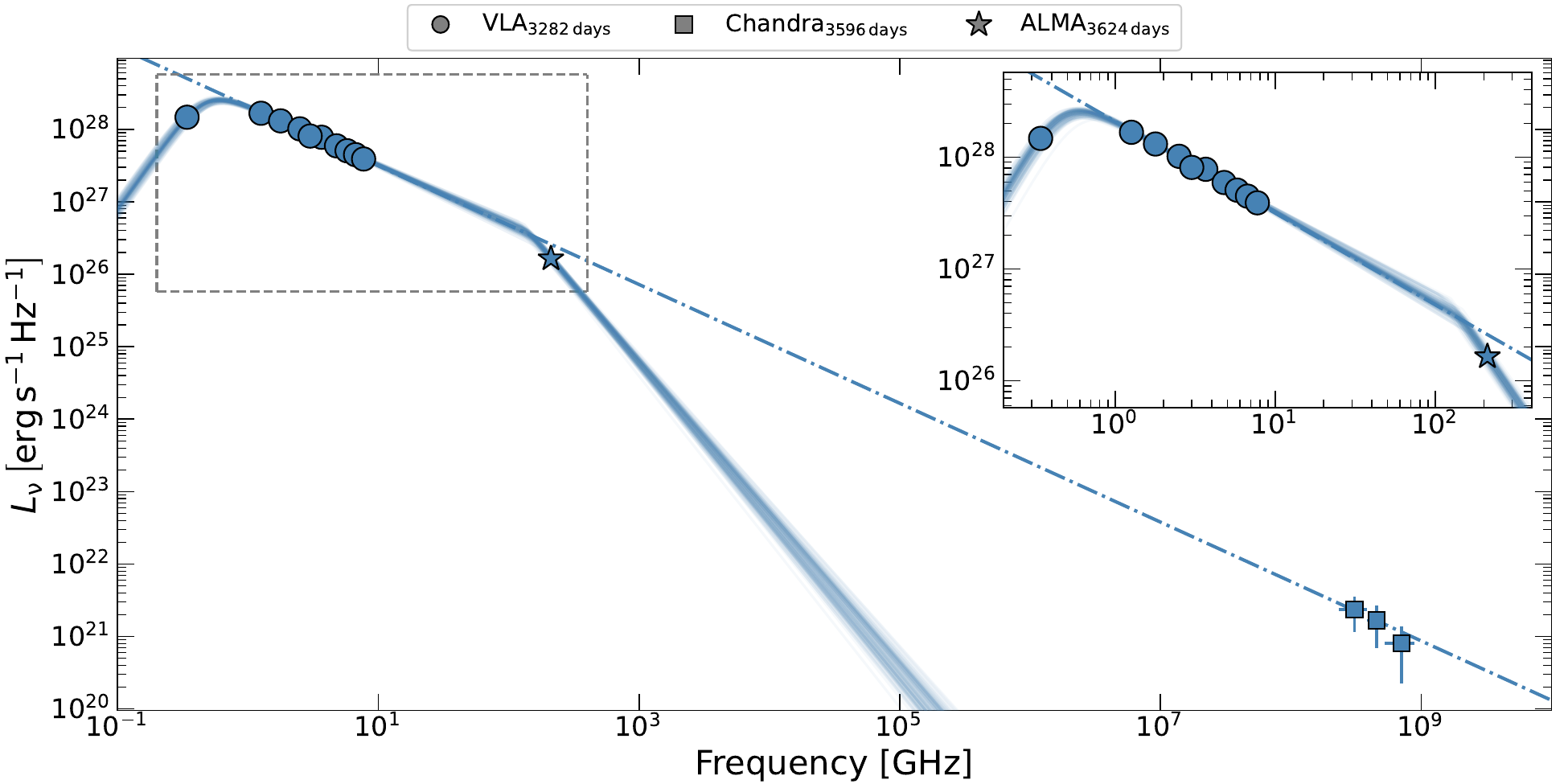}
    \caption{The near-simultaneous SEDs of SN\,2012ap in the time interval $\delta t \sim$ 12--4561\,d. The top left panel shows the early observations by \citet{Chakraborti_2015} from $\delta t \sim $12--38\,d, while the top right panel shows the late-time radio emission reported here. We fit the SEDs with the broken power-laws function described in Eq.~\ref{eq: bpl_function}. The bottom panel presents the broadband radio-to-X-ray SED. Here, we fit the radio-to-mm emission with a double broken power-law function (VLA+ALMA; solid lines). In addition, we show the extrapolation of the middle segment power-law to the X-ray bands (dotted-dashed line). This extrapolation naturally accounts for the observed X-ray emission (which was not fitted). However, it over-estimates the emission in mm-band, which is easily explained by a break frequency introduced at $\nu_{\rm c} = 140 \pm 20$ GHz. See detailed discussion in \S\ref{subsec: density_enhancement}. }
    \label{fig: SN2012ap_radio}
\end{figure*}

Both the radio and the X-ray emission from SN\,2012ap exhibit two distinct phases. Initially, there was no X-ray emission detected from this SN \citep{Margutti_2014}. Then, our late-time Chandra observations revealed a bright source ($L_{x} \approx 3 \times 10^{39}\, {\rm erg \, s^{-1}}$) that remains constant in time at $\delta t \gtrsim 9.8$\,years (see \S~\ref{SubSec:Chandra}). In comparison, the early radio emission component, observed at $\delta t = 12 - 38$\,d (first reported and analyzed by \citealt{Chakraborti_2015}; see also Fig.~\ref{fig: SN2012ap_peaks} and the top left panel of Fig.~\ref{fig: SN2012ap_radio}), revealed broken-power-law SEDs with a declining peak flux density, $F_{\rm p} \sim t^{-0.4}$, peak frequency, $\nu_{\rm p} \sim t^{-0.7}$, and an observed peak luminosity density of $L_{\rm \nu} \sim 10^{28} \, \rm erg \, s^{-1} \, Hz^{-1}$. A second radio component, brighter by a factor of $\sim 2$ than the early component, is observed at $\delta t = 3282 - 4651$\,d (first reported in the study of a VLASS sample by \citealt{Stroh_2021}; see the top right panel of Fig.~\ref{fig: SN2012ap_radio}). This late-time emission is mostly optically thin in the GHz bands with a peak at $\nu_{\rm p} \sim 0.4$ GHz. Interestingly, this emission component does not seem to evolve significantly over a timescale of more than three years (see Fig.~\ref{fig: SN2012ap_peaks} for the temporal evolution of the radio spectral peaks). 

Extrapolating the late-time radio SED from the GHz bands to the X-ray band results in a perfect match of the X-ray flux density (see the bottom panel of Fig.~\ref{fig: SN2012ap_radio}). Furthermore, the similarity between the radio and X-ray emission extends to their lack of temporal evolution. These radio-to-X-ray properties might suggest that the X-ray emission is just the tail of the synchrotron SED seen in the radio. However, an ALMA observation with a central frequency of $211$ GHz at $\delta t = 3624$\,d analyzed together with all the late-time broadband radio SEDs at $\delta t \geq 3282$\,d point to the presence of a spectral break at frequency $\nu < 211$ GHz (see bottom panel of Fig.~\ref{fig: SN2012ap_radio}). Therefore, the X-ray emission can be a result of a different emission mechanism (e.g., bremsstrahlung emission from either the reverse or forward shock), and the matching of the synchrotron tail to the X-ray emission may be a coincidence. In the following analysis of the broadband emission (radio and X-ray) we assume that the spectral break observed in the mm-bands is due to a break frequency at $\nu < 211$ GHz, and associate the X-ray emission with thermal emission. We justify this choice by showing that requiring the cooling break frequency to be above the X-ray band implies highly nonphysical parameters (see Appendix \S\ref{subsec: cool_disc}).

Radio emission from SNe is typically associated with synchrotron emission that originates from the interaction between the SN ejecta and the ambient medium. This interaction drives a shockwave, and at the shock front magnetic fields are compressed and generated. In addition, electrons are accelerated to relativistic velocities and gain a power-law distribution of Lorentz factors, $N \left( \gamma \right) \, {\rm d} \gamma \propto \gamma^{-p} \, {\rm d} \gamma; \, \gamma \geq \gamma_{\rm m}$, where $\gamma_{\rm m}$ is the Lorentz factor of the lowest energy electrons \citep{Sari_1998, Chevalier_1998}. The spectral shape of the broadband SEDs is dictated by the order of the minimum Lorentz factor, $\gamma_{\rm m}$, the Lorentz factor above which the electrons are cooled and lose significant energy to radiation, $\gamma_{\rm c}$, and, $\gamma_{\rm sa}$, below which the emission from the electrons is self-absorbed \citep{Granot_2002}. In the slow-cooling regime ($\gamma_{\rm m} < \gamma_{\rm c}$), and for $\nu_{\rm m} < \nu_{\rm sa}$ (here we defined $\nu_{i}$ as the frequency that corresponds to $\gamma_{i}$), the optically thick emission is $F_{\rm \nu} \sim \nu^{5/2}$ (where $F_{\rm \nu}$ is the flux density). The optically thin emission goes as $F_{\rm \nu} \sim \nu^{-p/2}$ for $\nu_{\rm c} < \nu_{\rm sa}$. On the other hand, if $\nu_{\rm sa} < \nu_{\rm c}$ another break in the SED is observed, and $F_{\rm \nu} \sim \nu^{-\left( p-1 \right)/2}$ for $\nu < \nu_{\rm c}$, and $F_{\rm \nu} \sim \nu^{-p/2}$ for $\nu > \nu_{\rm c}$.

A typical assumption in this SN-CSM interaction model is that a fraction, $\epsilon_{\rm e}$, of the post-shock energy density, $u_{\rm ps} \propto \rho_{\rm csm} v_{\rm sh}^2$ (where $\rho_{\rm csm}$ is the density of the CSM, and $v_{\rm sh}$ is the shock velocity), is deposited in the relativistic electrons. Another fraction, $\epsilon_{\rm B}$, is transferred to the energy density of the magnetic field, $u_{\rm B} \propto B^2$, meaning that $B \propto \epsilon_{\rm B}^{1/2} \rho_{\rm csm}^{1/2} v_{\rm sh}$. The synchrotron self-absorbed emission evolves as $F_{\rm \nu} \sim R^2 B^{-1/2}$ in the optically thick regime, and $F_{\rm \nu} \sim R^3 B^{\left( p+5 \right)/2}$ in the optically thin regime \citep{Chevalier_1998}. These power-laws imply that a monotonic evolution of the radio emission is expected unless there is a change in the density structure and/or in the hydrodynamic evolution of the shockwave.

In this simple SN-CSM interaction scenario, the early radio emission from SN\,2012ap is easily explained with a mildly-relativistic outflow interacting with a single power-law density profile \citep{Chakraborti_2015}. However, the observed late-time rebrightening of the radio emission does not match the extrapolation of the early radio component and therefore it might signal of a density increase at large radii. This would imply that the progenitor massive star experienced different phases of mass-loss during the last stages of its evolution. On the other hand, the early emission and the late-time radio rebrightening can be a result of an off-axis jet+cocoon system, in which the early emission would be due to a mildly relativistic outflow with a broad angular structure (the cocoon), and the late time component would be a result of the relativistic narrow jet entering our line of sight. If confirmed, this scenario would imply that the lack of a long GRB detection associated with SN\,2012ap and the low energy inferred for the fastest outflow at early times is due to geometry effects and not an intrinsically weak engine-driven explosion. In the following sub-sections we explore these two scenarios.

\subsection{A density enhancement at $R \gtrsim 10^{17} \, \rm cm$}
\label{subsec: density_enhancement}

We first explore the scenario in which the observed radio and X-ray rebrightening is due to a density increase at large radii. We present the analysis of the radio emission in \S~\ref{subsub:radioFS}, and of the X-ray emission in \S~\ref{subsub:XraysRS}. However, we emphasize that this analysis was done in parallel and the set of microphysical parameters in \S~\ref{subsub:radioFS} were chosen to best match the analysis in \S~\ref{subsub:XraysRS}.

\subsubsection{Radio emission from the forward shock in the CSM}
\label{subsub:radioFS}

We adopt the synchrotron self-absorption (SSA) analysis presented in \cite{Chevalier_1998} to estimate the density profile around SN\,2012ap. This analysis assumes that the observed peak in the broadband SED is at $\nu_{\rm sa}$, and that $\nu_{\rm m} < \nu_{\rm sa} < \nu_{\rm c}$ (we check for self-consistency of these frequencies within the GHz bands in this analysis). The radius of the emitting shell is
\begin{align}
    \label{eq: radius_chevalier}
    R = \left[ \frac{6 c_{\rm 6}^{p+5} F_{\rm \nu_{\rm sa}}^{p+6} d_{\rm l}^{2p+12}}{\alpha f (p-2) \pi^{p+5} c_{\rm 5}^{p+6} E_{\rm l}^{p-2}} \right]^{1/(2p+13)} \left( \frac{\nu_{\rm sa}}{2c_{\rm 1}} \right)^{-1}
\end{align}
and the magnetic field strength is
\begin{align}
    \label{eq: bfield_chevalier}
    B = \left[ \frac{36 \pi^{3} c_{\rm 5}}{\alpha^2 f^2 (p-2)^2 c_{\rm 6}^3 E_{\rm l}^{2(p-2)} F_{\rm \nu_{\rm sa}} d_{\rm l}^2} \right]^{2/(2p+13)} \left( \frac{\nu_{\rm sa}}{2c_{\rm 1}} \right)
\end{align}
where $F_{\rm \nu_{\rm sa}}$ is the peak flux density at the intersection between the optically thick and thin regimes of the synchrotron self-absorbed SED; $d_{l}$ is the luminosity distance to the SN; $f$ is the emission volume filling factor; $\alpha \equiv \epsilon_{\rm e}/\epsilon_{\rm B}$; $c_{\rm 1}$, $c_{\rm 5}$, and $c_{\rm 6}$ are constants provided in \cite{Pacholczyk_1970}; and $E_{\rm l}$ is the electron's rest-mass energy. Under the assumption of a strong shock, the CSM density is given by
\begin{align}
\label{eq: density_bfield}
    \rho_{\rm csm} = \frac{1}{9 \pi} \epsilon_{\rm B}^{-1} B^2 v_{\rm sh}^{-2}.
\end{align}
Here we assumed that the post-shock energy density is $9/8 \rho_{\rm{csm}} v_{\rm{sh}}^2$ \citep{DeMarchi_2022, Sfaradi_2025_tvd}. Note that this description implicitly assumes that the shock is in the deep-Newtonian regime (see \citealt{Minhajur_2026} for a more detailed discussion in the deep-Newtonian regime in sub-relativistic outflows).

To obtain $F_{\rm \nu_{\rm sa}}$, $\nu_{\rm sa}$, and $p$, we adopt the broken power-law function (see Eq. 1 in \citealt{Granot_2002})
\begin{align}
    \label{eq: bpl_function}
    F_{\rm \nu} = F_{\rm \nu_{\rm sa}} \left[ \left( \frac{\nu}{\nu_{\rm sa}} \right)^{-s\frac{5}{2}} + \left( \frac{\nu}{\nu_{\rm sa}} \right)^{s\left( p-1 \right)/2} \right]^{-1/s}
\end{align}
where $s = 1.25 - 0.18p$ is the smoothing parameter (which is described in \citealt{Granot_2002}). We use implement MCMC using \texttt{emcee} \citep{Foreman_Mackey_2013} to determine the best-fit parameters and infer their posterior distributions using flat priors. We use 200 walkers with $20,000$ steps for each chain and discard the first $1000$ steps for burn-in. We test for convergence by running the chains for at least $50\tau$ where $\tau$ is the autocorrelation time. GMRT and VLASS observations at $\delta t = 3844$ and $4039$\,d, respectively, are combined to achieve a broadband SED. We fit for the power-law index of the electrons, $p$, at $\delta t = 18, 3282, 3844-4039$, and $4651$\,d, and apply the power-law index inferred at $\delta t = 18$\,d to the broadband SEDs at $\delta t = 12, 27$, and $38$\,d. We report the results of this analysis in Table~\ref{tab: bpl_fits_parameters} and plot best-fitting models in Fig.~\ref{fig: SN2012ap_radio}. To obtain the cooling break frequency, $\nu_{\rm c}$, we combine the broadband VLA SED at $\delta t = 3282$\,d with the ALMA observation at $\delta t = 3624$\,d. We fit this VLA+ALMA SED with the broken power-law model in Eq.~\ref{eq: bpl_function}, and introduce a sharp break at $\nu_{\rm c}$ above which $F_{\rm \nu} \sim \nu^{-p/2}$ (and use \texttt{emcee} with the same priors and parameters described above). This analysis results with $\nu_{\rm c} = 140 \pm 20$ GHz (see bottom panel of Fig.~\ref{fig: SN2012ap_radio}).

We derive the physical parameters of the shock and its environment in light of the SN-CSM interaction model using these spectral peaks\footnote{\cite{Chakraborti_2015} performed this analysis for the early emission but used the observed peak flux density and not the flux at the intersection between the optically thin and thick regime. For self-consistency, we re-derive the physical parameters for these early epochs.}. First, we assume a spherical outflow with an emission filling factor of $f=0.5$. For the energy partition parameters, assuming that the break frequency in the ALMA band is due to synchrotron cooling, we find that the results are consistent with deviation from equipartition ($\epsilon_{\rm e}/\epsilon_{\rm B} \approx 0.1$) for $\delta t \geq 3282$\,d and we use this ratio for the analysis of the late-time emission. For the early time emission ($\delta t \leq 38$\,d), we start by assuming equipartition and that $\epsilon_{\rm e} = \epsilon_{\rm B} = 0.1$. We note that while this assumption is common in the SN literature, it is not justified because we cannot break the degeneracy between the microphysical parameters at these early times. Therefore, we also derive the results using $\epsilon_{\rm e}/\epsilon_{\rm B} \approx 0.1$ (similar to the analysis of the late-time SEDs).

\begin{figure*}[ht]
\centering
\includegraphics[width=\linewidth]{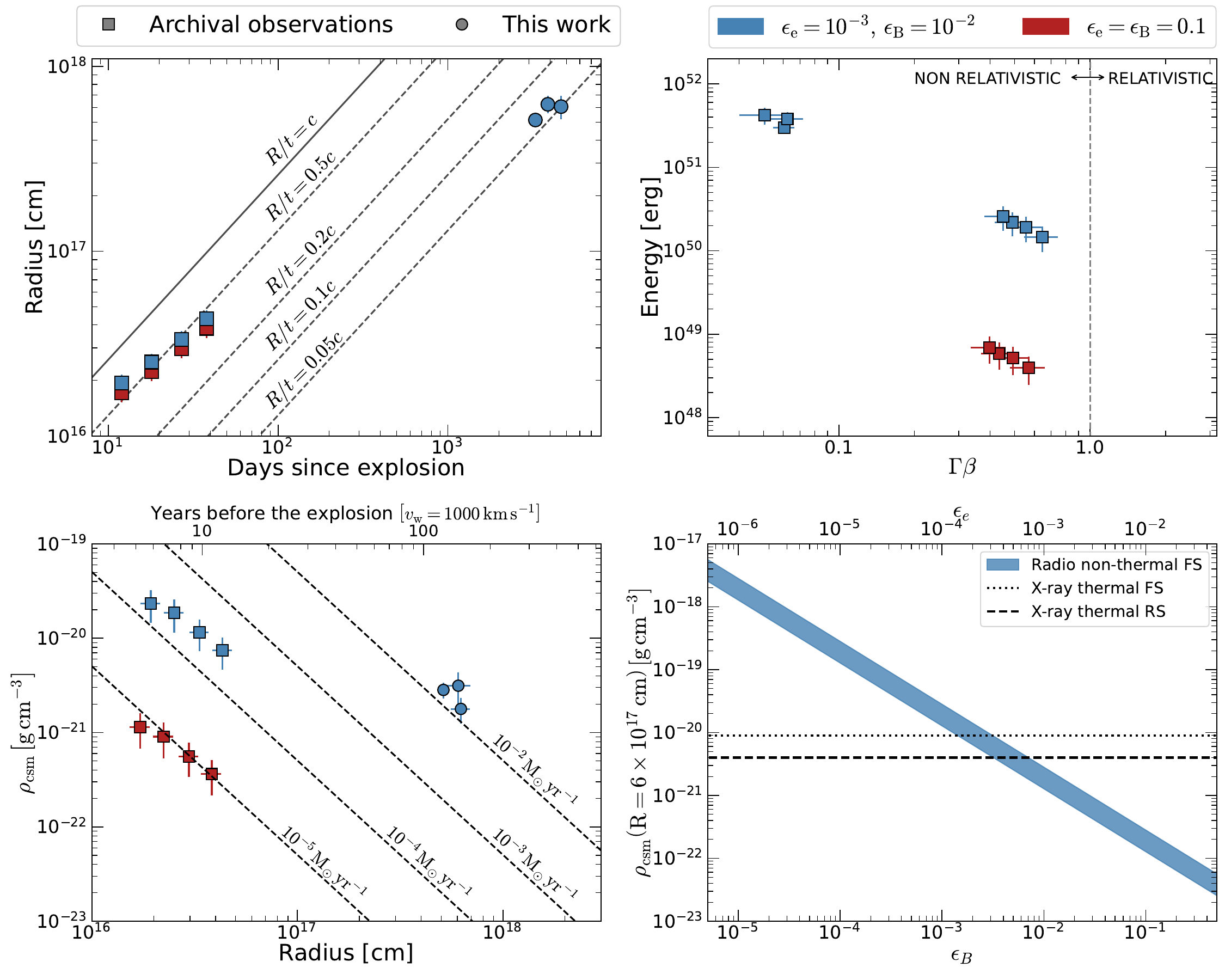}
\caption{Physical parameters inferred from the fits of the individual SEDs and the SSA analysis (see \S\ref{subsub:radioFS}). The top left panel shows the temporal evolution of the radius, the top right panel presents the energy-velocity profile. The bottom left panel is for density profile, also plotted for reference are line of equal mass-loss rate assuming a wind velocity of $1000 \, \rm km \, s^{-1}$. In the bottom right panel we present the density of the CSM at $R = 6 \times 10^{17}$\,cm derived from the late-time radio SEDs for different values of $\epsilon_{\rm B}$ while keeping a constant ratio of $\epsilon_{\rm e}/\epsilon_{\rm B} = 0.1$ (blue stripe). We also plot the density at $R = 6 \times 10^{17}$\,cm derived from the analysis of the X-ray emission using a forward shock (FS) model (dotted line) and a reverse shock (RS) model (dashed line). This is discussed in detaild in \S\ref{subsub:XraysRS}. The analysis presented in the bottom right panel was used to determine $\epsilon_{\rm B} \simeq 10^{-2}$ and $\epsilon_{\rm e} \simeq 10^{-3}$ in our SSA analysis (the results of this are shown in the other three panels).
\label{fig: equipartition}}
\end{figure*}

We start by estimating the radius of the outflow at different times since it depends only on the ratio of microphysical parameters ($\alpha = \epsilon_{\rm e} / \epsilon_{\rm B}$) and not the actual values (see Eq.~\ref{eq: radius_chevalier}). We report these values in Table~\ref{tab: equipartition_parameters} and plot them in the upper left panel of Fig.~\ref{fig: equipartition}. The temporal evolution of the radius implies a decelerating, mildly-relativistic, shock wave with $R \sim t^{0.7}$ and $\Gamma \beta \approx 0.4-0.7$ up to $\delta t = 38$\,d, where we estimate the mean velocity, $\beta = R/ct$, and $\Gamma \equiv 1/\sqrt{1-\beta^2}$. The radius at late-times ($\delta t \geq 3282$\,d) is consistent with $R \approx 5-6 \times 10^{17}$\,cm, sub-relativistic velocities ($\Gamma \beta \approx 0.05-0.06$), and no significant temporal evolution. 

In order to estimate the energy of the outflow and the density of the CSM we need to assume the values of $\epsilon_{\rm e}$ and $\epsilon_{\rm B}$. We can break the degeneracy in these parameters for the radio SEDs obtained at late-times by estimating the density at $R \simeq 6 \times 10^{17}$\,cm assuming that the X-ray emission is due to bremsstrahlung emission from either the shocked material behind the FS or the RS (see detailed discussion in \S~\ref{subsub:XraysRS}). From this analysis, we find $\epsilon_{\rm e} \simeq 10^{-3}$ and $\epsilon_{\rm B} \simeq 10^{-2}$ for $\delta t \geq 3282$\,d (see bottom right panel of Fig.~\ref{fig: equipartition}). We next use these values of $\epsilon_{\rm e}$ and $\epsilon_{\rm B}$ to estimate the kinetic energy of the shock and the CSM density, assuming spherical shockwave with an emitting filling factor, $f=0.5$ (see Fig.~\ref{fig: equipartition} and Table~\ref{tab: equipartition_parameters} in Appendix \ref{sec: tables}). Note that we report the physical parameters of the shock at early times by assuming both equipartition ($\epsilon_{\rm e} = \epsilon_{\rm B} = 0.1$) and $\epsilon_{\rm e} \simeq 10^{-3}$ and $\epsilon_{\rm B} \simeq 10^{-2}$ (similar to the analysis of the late-time observations). 

The structure of the CSM density changes around $5 \times 10^{16} - 5 \times 10^{17} \, \rm cm$ (see the bottom left panel of Fig.~\ref{fig: equipartition}). Initially, $\rho_{\rm csm} \sim r^{-2}$, consistent with constant mass-loss rate in steady winds. However, the observed density at $R \approx 6 \times 10^{17} \rm \, cm$ is higher by two or three orders of magnitude (depending on the choice of $\epsilon_{\rm e}$ and $\epsilon_{\rm B}$ at early times) than the extrapolation of this profile. Assuming that the CSM was deposited in mass-loss by winds of $1000 \, \rm km \, s^{-1}$, for $\epsilon_{\rm e} = \epsilon_{\rm B} = 0.1$, we infer a mass-loss rate of $\dot{M} \approx 10^{-5} \, \rm M_{\odot} \, yr^{-1}$ during the last $\sim15$ years of stellar evolution. This assumes that the wind is freely expanding; however, since it is expanding to a dense medium, some deceleration is warranted (we do not account for it as we do not know the exact density profile of the CSM). At earlier stages of the stellar evolution, at least $\sim 15$ years before the stellar explosion, the mass-loss was significantly higher, with $\dot{M} \approx 10^{-2} \, \rm M_{\odot} \, yr^{-1}$ , for a constant wind velocity. It is likely that the change in density arises from a change in the wind velocity, from $\sim 10 \, \rm km \, s^{-1}$ to $\sim 1000 \, \rm km \, s^{-1}$ in the last stages of stellar evolution, possibly due to a transition from the red supergiant to the Wolf-Rayet phase.

In this scenario, the total mass in the CSM is large. Assuming that the change occurs around $R_{\rm br} \sim 5 \times 10^{17} \, \rm cm$, the total mass in the CSM up to $6 \times 10^{17}$\,cm is $\sim 1 \, M_{\odot}$. This can explain the sharp deceleration of the shock wave from $\approx 0.5c$ to $\approx 0.05c$, however, it requires the deposition of a large fraction of the mass of an H-stripped progenitor star timed with its core-collapse. In addition, the energy required for this sub-relativistic outflow is $E_{\rm k} \approx 3-4 \times 10^{51} \, {\rm erg}$, about $15\%$ of the kinetic energy inferred at early times, and more energetic than any radio emitting sub-relativistic outflow observed so far. However, it is important to note that most radio emitting sub-relativistic outflows are detected at early times when most of the energy is carried out by the slow moving ejecta and the radio emission is due to the fastest ejecta.

\subsubsection{X-rays from thermal emission of the FS or  RS of the SN}
\label{subsub:XraysRS}

The extrapolation of the synchrotron spectrum of \S\ref{subsub:radioFS} that fits the radio emission to $\nu$$>$$10^{17}\,\rm{Hz}$ (with the assumption of a cooling break at $\sim 140$\,GHz) significantly under-predicts the X-ray observations. Furthermore, we show in Appendix \ref{subsec: cool_disc} that if we consider the X-rays to be an extension of the synchrotron FS emission seen in the radio we infer nonphysical michrophysical parameters ($\epsilon_{\rm e} / \epsilon_{\rm B} \gtrsim 10^{8}$). Therefore, we exclude synchrotron FS emission as the origin of the X-rays. However, in SN shocks expanding in high-density media it is often the case that the X-ray band can be dominated by thermal bremsstrahlung emission from shocked material behind the FS or the RS (see e.g., SNe 2014C or 2013ef; \citealt{Brethauer22, Kamble_2016}).

The density of the X-ray \emph{emitting} region $\rho_{\rm{X}}$ can be derived from the constraints on the normalization of the bremsstrahlung model ($\rm{Norm}\equiv\frac{3.02\times 10^{-15}}{4\pi d^2  }\int  n_e n_I dV$, where $n_e=\rho_X/\mu_e m_p$ and $n_I=\rho_X/\mu_I m_p$)  and depends on (i) the chemical composition ($\mu_e$ and $\mu_I$) and (ii) the emitting volume $V$. The inferred CSM density $\rho_{\rm{csm}}$ further depends on  (iii) the FS vs. RS nature of the emitting region, and on (iv) the ultra-relativistic or non-relativistic nature of the shock. We will consider Solar ($\odot$) and Ibc ejecta-like  ($\star$) compositions, and non-relativistic shocks (relativistic shocks are explored in \S\ref{subsec: jet_cocoon}). We parametrize the emitting volume as some fraction $f_X$ of a shell between $R_{\rm in}$ and $R_{\rm out}$ such that $R_{\rm in} \equiv aR_{\rm out}$ and $V=\frac{4\pi}{3}R_{\rm out}^3(1-a^3)f_X$, from which it follows that filling factor of the radio model is $f=(1-a^3)f_X$. It follows that: \begin{equation}
\bar{{\rho_X}}\propto {\rm Norm^{1/2}}\frac{m_p(\mu_e\mu_I)^{1/2}}{V^{1/2}} \propto  \frac{m_p(\mu_e\mu_I)^{1/2}}{R_{\rm out}^{3/2}(1-a^3)^{1/2}f_X^{1/2}},
\end{equation}
\noindent where $\rho_X$ is an average density of the emitting region ($\bar{{\rho_X}}\equiv (<\rho_X^2>_V)^{1/2}$). Since $\frac{\sqrt{\mu_e\mu_I}|_{\odot}}{\sqrt{\mu_e\mu_I}|_{\star}}\approx0.37$, the assumption of Solar- vs. ejecta-like composition leads to $\hat{\rho_X}|_{\odot}\approx 0.37 \hat{\rho_X}|_{\star}$ everything else being equal.

\emph{FS emitting region:} In this case, for a strong shock and gas with adiabatic index $=5/3$, $\rho_{\rm{csm}}\approx \frac{\hat{\rho_X}}{4}$. The X-ray emitting region is the same as the radio emitting region for which  $f=(1-a^3)f_X=0.5$ (from \S\ref{subsub:radioFS}). For Solar composition we infer $\rho_{\rm{csm}}(R_{\rm out})=9\times 10^{-21}\Big (\frac{R_{\rm out}}{6\times 10^{17}\rm{cm}}\Big )^{-3/2}\rm{g\,cm^{-3}}$, implying agreement with the radio inferences of \S\ref{subsub:radioFS} (see Fig.~\ref{fig: equipartition}) for $\epsilon_B\lesssim 10^{-2}$ and $\epsilon_e\lesssim 10^{-3}$ (where the ratio $\epsilon_e/\epsilon_B=0.1$ has been chosen to satisfy the $\nu_c$ constraint implied by the ALMA observations). For these values the internal energy is $\sim 3 \times 10^{51}$\,erg.

\emph{RS emitting region:} Several years after the explosion it is plausible that the X-ray emission from the SN RS can escape and reach the observer even if a cold dense shell had formed between the RS and the FS (e.g., \citealt{Chevalier17}). In this case the radio and the X-ray radiation originate from different emitting volumes. During the self-similar phase, requiring the continuity of the pressure at the contact discontinuity and pressure balance, the density ratio of forward shocked to reverse shocked material is $\frac{\rho_{\rm RS}}{\rho_{\rm FS}}\approx \Big (\frac{n-3}{3-s}\Big )^2 \Big ( \frac{R_{\rm RS}}{R_{\rm FS}}\Big)^2$ (e.g., \citealt{Chevalier82self}). In our scenario $\rho_{\rm RS}\equiv \hat{\rho_X}$, $\rho_{\rm{csm}}\approx \rho_{FS}/4$ and we assume ejecta-like composition for the emitting region, as appropriate for reverse-shocked material. We infer
$\rho_{\rm{csm}}(R_{out})=4\times 10^{-21}\Big (\frac{R_{out}}{6\times 10^{17}\rm{cm}}\Big )^{-3/2}\rm{g\,cm^{-3}}$, 
where we have adopted $a=0.98$, $f_X=1$, and $\frac{\rm R_{RS}}{\rm R_{FS}}\approx 0.8$, as appropriate for $s$ in the range $\approx 1-2$, and $n\approx10$. An important caveat of this analysis is that at these late times the RS is likely to have reached the inner flatter portion of the ejecta profile and the self-similar solutions might not apply. Treating the results above as an order of magnitude estimate, and similar to our previous conclusions, we find agreement with the radio CSM densities  for $\epsilon_B\lesssim 10^{-2}$ for $\epsilon_e/\epsilon_B=0.1$, implying large internal energy coupled with the FS region ($E_{\rm k}\sim 3 \times 10^{51}$\,erg). We note however that in principle the RS does not have to share the same shock microphysics as the FS.

This analysis shows that the SN RS and FS might be contributing comparable amounts of X-ray emission at this epoch and can explain the observations.

\subsection{A jet + cocoon structure observed off-axis}
\label{subsec: jet_cocoon}

When viewed off-axis, relativistic jets are invoked as a mechanism to explain the emergence of late-time non-thermal emission across different type of transients (e.g., the Type Ic-BL SN PTF\,10tqv, \citealt{Scheroder_2025}; the TDE\,2018hyz, \citealt{Matsumoto_2023, Sfaradi_2024}; GW\,170817, \citealt{Margutti_2021} and the references therein). In this scenario, the relativistic jet is launched away from our line of sight, and the emission is beamed relativistically, with a beaming angle of $\Gamma^{-1}$, resulting in lack of non-thermal emission early on. Then, as the jet decelerates the bulk Lorentz factor decreases, the beaming effect is reduced, and the emission enters our line of sight, resulting in the late-time brightening. The emission is expected to peak roughly when the beaming cone is wide enough to enter our line of sight, this translates to $\Gamma \approx \theta_{\rm obs}^{-1}$ (e.g., \citealt{Granot_2002b, Margutti_2017}). \cite{Granot_2002, Granot_2003} showed that a relativistic shockwave traveling through the ISM will evolve as $\Gamma \left( t \right) = 6.68 \left( \frac{E_{ {\rm k}, \, iso, \, 52}}{n_{\rm {\rm 0}}} \right)^{1/8} t_{\rm days}^{-3/8}$, where $E_{ {\rm k}, \, iso, \, 52}$ is the isotropic equivalent energy is units of $10^{52} \, \rm erg$, $n_{\rm 0}$ is the number density of the ISM in $\rm cm^{-3}$, and $t_{\rm days}$ is the time in days since explosion in days (see also the discussion in \citealt{Margutti_2017}). Therefore, at the time of the peak, $t_{\rm pk}$, the viewing angle is $\theta_{\rm obs} = 0.15 \left( \frac{E_{ {\rm k}, \, iso, \, 52}}{n_{\rm {\rm 0}}} \right)^{-1/8} t_{\rm pk, \, days}^{3/8}$. Overall, the late-time radio emission from SN\,2012ap is flat and it is possible that it is around peak at $\sim 3000-4000$\,d after stellar explosion. Assuming an ISM density of $n_{\rm 0} \sim 1 \, \rm cm^{-3}$ and that the collimated corrected energy is $E_{k} \approx 0.5 \theta^{2}_{0} E_{k, \, \rm iso}$ we get $\theta_{\rm 0} \approx 5.3^{\circ} \times \left[ \frac{\theta_{\rm obs}}{90^{\circ}}\right]^4 \left[ \frac{E_{k}}{2 \times 10^{52} \, \rm erg} \right]^{1/2}$. This motivates us to search for relativistic jets viewed extremely off-axis ($\sim 60^{\circ}-90^{\circ}$) since opening angles smaller than a few degrees are not expected and a jet energy larger than a ${\rm few} \times 10^{52} \, \rm erg$ is not expected as the energy of the SN ejecta inferred from early observations was $\sim 2 \times 10^{52} \, \rm erg$ \citep{Milisavljevic_2015}.

Motivated by this analysis, and following the outflow structure observed in GRBs \citep{Woosley_2006}, we explore here a two component outflow structure of a collimated relativistic jet surrounded by a wide angle, mildly-relativistic, outflow (i.e., cocoon). To model the emission, we use \texttt{VegasAfterglow} \citep{Wang_2026}, a framework that self-consistently solves forward and reverse shock dynamics and calculates synchrotron emission. For simplicity, we assume a non-spreading top-hat jet structure for each component, and a CSM-like density structure, $\rho \sim r^{-k}$, that transitions to an ISM density structure (i.e., constant density profile). We note here that we explored models of spreading jets using \texttt{VegasAfterglow}; however, these models failed to describe the slow evolution of the light curve at late times.

This model has 16 parameters, and there are naturally many degeneracies. For example, the cocoon emission is consistent with both spherical and collimated outflows. Therefore, we cannot constrain its opening angle and fix it to $\theta_{0, \, c} = 60^{\circ}$. We note that this is somewhat arbitrary and that a wider opening angle will result in lower energetics for the cocoon. In addition, the energy partition parameters of the cocoon are not constrained and we set them to $\epsilon_{e, \, c} = \epsilon_{B, \, c} = 0.1$. Finally, we also fix the density of the ISM to be $n_{\rm ISM} = 1 \, {\rm cm^{-3}}$. We use \texttt{emcee} \citep{Foreman_Mackey_2013} to fit our model with 12 free parameters, and use flat priors. We summarize our model parameters, their definitions, their priors, and the best-fitting values in Table~\ref{tab: jet_cocoon_model_parameters} in appendix \ref{sec: tables}.

\begin{figure*}[ht]
\centering
\includegraphics[width=0.49\linewidth]{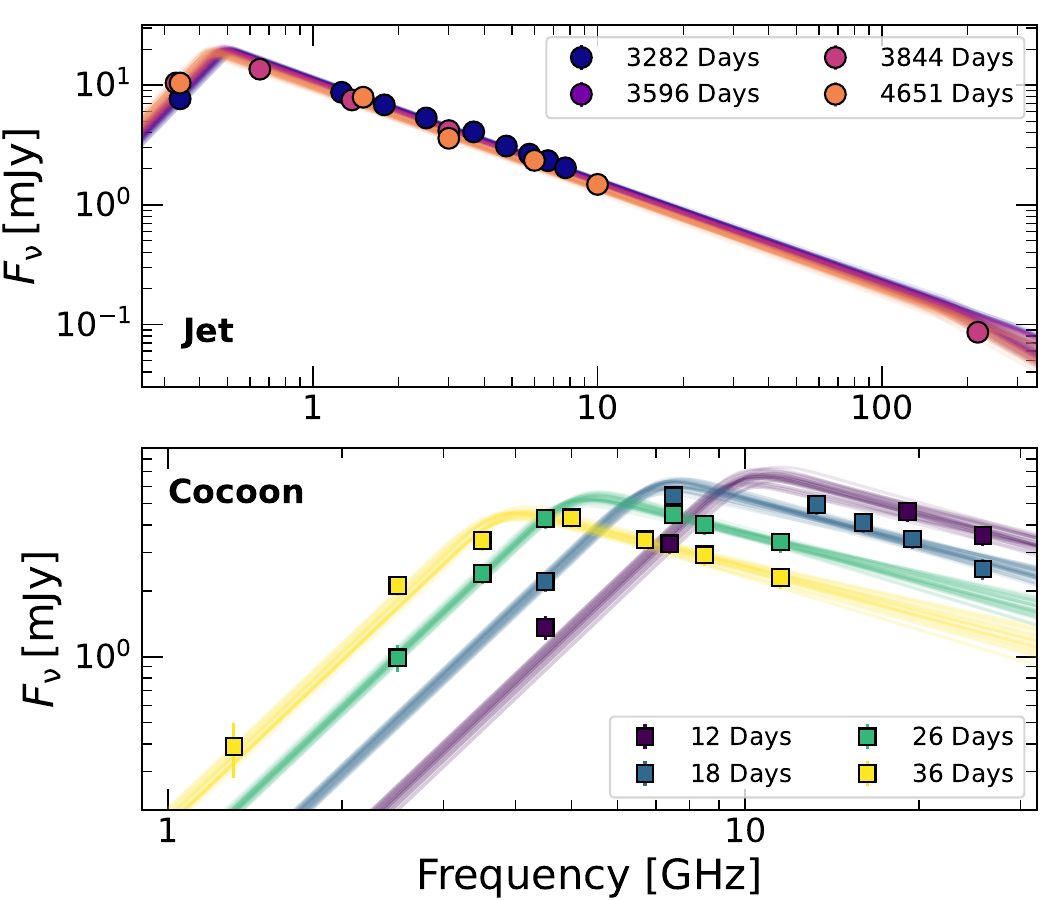}
\includegraphics[width=0.5\linewidth]{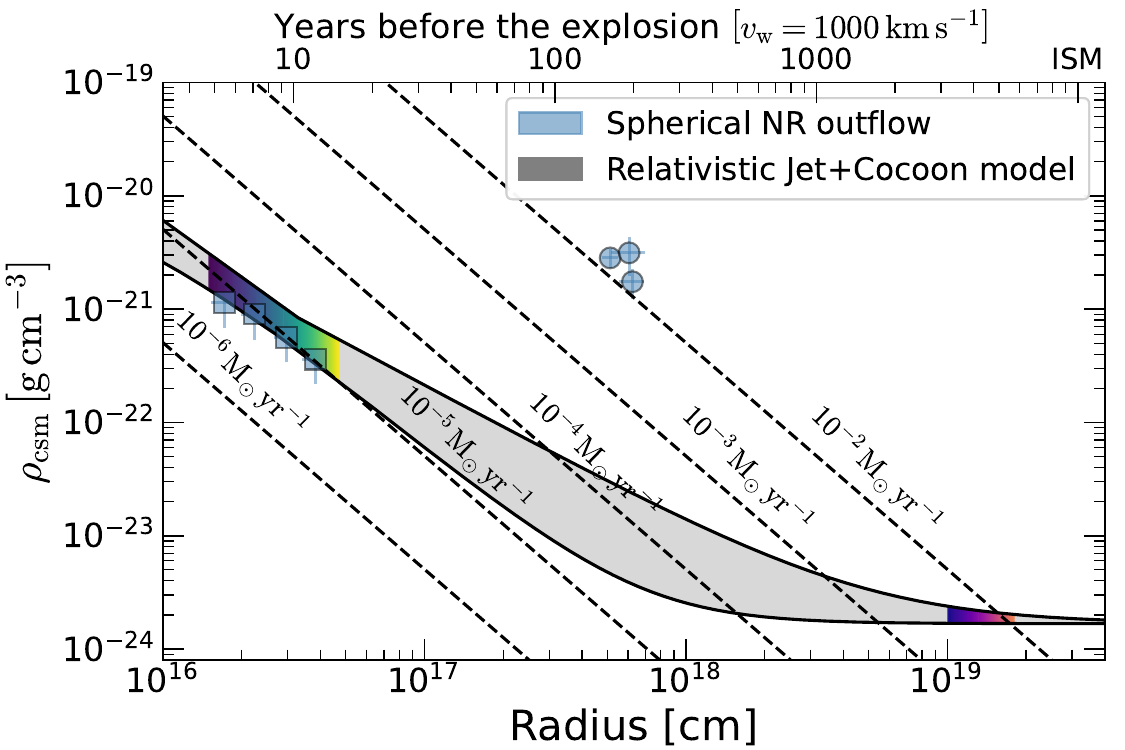}
\caption{Best-fitting results for the jet+cocoon model using \texttt{VegasAfterglow} code. The top left panel is for the late-time radio emission observed with the VLA, GMRT, and ALMA modeled with the jet component. The bottom left panel is for the early time radio emission observed with the VLA and modeled with the wide-angle outflow (i.e., the cocoon). The lines are drawn from the posterior distributions shown Fig.~\ref{fig: jet_cocoon_corner} in the appendix. The right panel presents the density profile inferred from this modeling (shaded gray region) together with the results from the analysis in \S~\ref{subsub:radioFS} (see Fig.~\ref{fig: equipartition}). The colored regions within the gray region are for the radii in which we have radio observations. We summarize the parameters of this model in Table~\ref{tab: jet_cocoon_model_parameters}.
\label{fig: jet_cocoon_fit}}
\end{figure*}

We find that the entire radio emission, spanning $12-4651$\,d after explosion, can be explained with our jet+cocoon model. Fig.~\ref{fig: jet_cocoon_fit} shows the broadband SEDs with our best-fitting model. The collimated corrected energy of the relativistic jet is consistent with $E_{k, \, True, \, j} \gtrsim 10^{52} \, {\rm erg}$ (note that it is bound by the upper limit of our prior, which is motivated by the energetics argument). The initial Lorentz factor is unbound and we find $\Gamma_{0, \, j} \geq 5$. The jet opening angle is $\theta_{0, \, j} \simeq 7.5^{\circ}$, and a large observing angle, $\theta_{\rm obs} \geq 80^{\circ}$ is needed. The jet opening angle and the viewing angle are degenerate, and, a smaller viewing angle requires a smaller opening angle (e.g., for $\theta_{\rm obs} = 70^{\circ}$ the opening angle is $\theta_{0, \, j} \approx 3^{\circ}$). We note that some of the assumptions and approximations implemented in \texttt{VegasAfterglow} may not fully capture the evolution of the jet at the very late-times we observe SN\,2012ap . As a result, the inferred physical parameters of the off-axis jet should be interpreted with appropriate caution. For the cocoon we find a moderate energy of $E_{k, \, iso, \, c} \sim 10^{50} \, {\rm erg}$ with $\Gamma_{0, \, c} \simeq 1.14$, equivalent to $\beta \simeq 0.5$. The density profile of the surrounding inner CSM is $\rho_{\rm csm} \sim r^{-1.55}$, shallower than a $r^{-2}$ profile (possibly due to a non-steady wind), and the transition to the ISM occurs at $\sim 10^{18} \, {\rm cm}$, meaning that by the time we first observe the relativistic jet, it is already traveling in the ISM. We provide corner plots of the posterior distributions in Fig.~\ref{fig: jet_cocoon_corner} in Appendix \ref{sec: corner_plots}.

While the jet+cocoon model presented above is in agreement with the observed radio data it cannot account for the late time X-ray emission (as a cooling break is predicted below the X-ray frequencies). Therefore, if the late-time radio emission is due to an off-axis relativistic jet, we need to introduce a different emission component to explain the X-rays. This component could be the thermal emission from either the RS or the FS introduced in \S~\ref{subsub:XraysRS}. However, here, the late-time radio emission is a result of the relativistic jet, and by invoking the RS or FS to explain the X-rays we risk over-producing the radio (as seen in \S~\ref{subsub:radioFS}). On the other hand, if we choose a lower value for $\epsilon_{\rm B}$ we can suppress the radio emission from the synchrotron FS emission. For example, choosing $\epsilon_{\rm B}=10^{-3}$ (instead of $\epsilon_{\rm B}=10^{-2}$) results in $F_{\rm \nu_{\rm sa}}\sim 1.4$\,mJy at $\nu_{\rm sa} \sim 0.13$\, GHz (for the same set of parameters used in \S~\ref{subsub:radioFS}). Thus, the radio emission from the FS in the CSM would be a factor of $\sim 50$ lower than the observed emission (and therefore negligible within the observation uncertainties). In conclusion, in the scenario where the radio emission is from a structure of a relativistic jet+cocoon observed off-axis, we can still explain the X-ray emission with thermal emission from either the RS or FS without over-producing the radio emission.

\section{Discussion}
\label{sec: disc}

The nature of the early radio emission from SN\,2012ap (up to $38$ days after explosion) is well understood as a result of a mildly relativistic outflow. However, the late-time radio emission ($\delta t >5$ years) is peculiar. The low GHz SEDs ($\leq 10$ GHz) feature rebrightening of the synchrotron emission that has barely evolved during $\approx 4$ years. In addition to the radio peculiarities, the X-ray emission also exhibits late-time brightening which is a remarkable match to the tail of the synchrotron SED observed in the low-GHz bands (see Fig.~\ref{fig: SN2012ap_radio}). On the other hand, this is in conflict with an observed cooling break at $\nu \approx 140$ GHz. Even without the observed cooling break below ALMA observation, we show that having the cooling break around the X-ray band requires non-physical microphysical parameters (see detailed derivation in Appendix \ref{subsec: cool_disc}). Finally, both the late-time optical spectra and the analysis of the X-rays as thermal emission point to large densities around the progenitor. However, the optical spectra lack the broad emission lines seen in other strongly interacting SNe at late-times.

In the following sections, we discuss the possible astrophysical scenarios and try to track down the origin of the late-time panchromatic emission.

\subsection{The origin of the late-time radio emission}
\label{subsec: origin_discussion}

Our modeling of the radio emission at late-times reveals either a sub-relativistic outflow interacting with two distinct density structures (requiring a density enhancement at large radii, $R_{\rm br} \approx {\rm few} \times 10^{17} \, {\rm cm}$), or a structure of a relativistic collimated jet and a wider outflow (the cocoon) observed at a large viewing angle. The CSM enhancement scenario provides a physical framework that naturally accounts for both the radio and X-ray emission with only one outflow. In addition, the optical emission suggests the presence of an unshocked dense CSM, enriched of CNO material and lacking Hydrogen, at large radii. However, it has two main drawbacks. The first is that the physical parameters we infer in the non-relativistic scenario require high energy in the shock and high mass in the CSM, presumably formed by the progenitor in the last stages of its evolution. The second drawback is that other SNe that show late-time interaction also exhibit broad or intermediate-width emission lines in their optical spectra, these are not present in the late-time optical spectra of SN\,2012ap (see \S\ref{sec: optical_observations}). 

On the other hand, while the relativistic jet+cocoon scenario requires two different outflows to explain the radio and X-ray emission, and while we cannot account for lateral spreading of the jet in our modeling, it provides a complete physical framework for both the early and late time radio emission, without the need to invoke sharp density change and high kinetic energy for a non-relativistic outflow. Finally, the existence of a relativistic off-axis jet provides a natural explanation for the unique properties of SN\,2012ap at early times and places it as part of the GRB population instead of as a weak engine-driven explosion (we discuss this in detail in \S~\ref{subsec: astro_context_discussion}).

Differentiating between these two scenarios requires additional measurements. For example, using polarization measurements is argued to be one way to distinguish between sub-relativistic outflows and relativistic jets \citep{Birenbaum_2024, Birenbaum_2026}. This is because the emission is expected to be more polarized around the time of the peak of a jetted outflow \citep{Granot_2002_polar, Gill_2018}. Another decisive way to differentiate between a non-relativistic outflow and a relativistic jet is by direct measurements of the position and size of the emitting region. Very Long Baseline Interferometry (VLBI) observations can be used to test exactly that \cite{Granot_2003}. If the late-time rebrightening is due to a non-relativistic spherical outflow at $6 \times 10^{17} \, {\rm cm}$, VLBI observations will result in a circular source with a size of $\sim 0.9 \, {\rm mas}$ which will be slightly resolved with the Very Long Baseline Array (VLBA) at $8.4$ GHz. On the other hand, while we only model the emission as arising from a non-spreading jet, any relativistic jet is likely to spread by the time we observe the radio emission. Therefore, the lack of a spreading jet solution might be due to the failure of the existing analytical frameworks to describe the jet spreading. If the late-time rebrightening is a result of a spreading relativistic jet, the size of the radio emitting region will be $\sim 2 \times 10^{19} \, {\rm cm}$ which will result in an angular size of $\sim 40 \, {\rm mas}$. Even if the jet significantly decelerated and the above mentioned size is over-estimated, VLBA observations will clearly distinguish between the two scenarios (relativistic jet vs\. non-relativistic spherical outflow) and help determine the physical origin of this outflow. We note that measuring the temporal evolution of the outflow size or the motion of the flux centroid would have been possible if an early VLBI observation of this source existed. Combining early and late VLBA observations could provide a direct measurement of the velocity of the outflow and therefore its nature.

As an alternative to the CSM enhancement and jet+cocoon scenarios, emission from PWN is sometimes invoked as a mechanism for slowly evolving radio emission associated with supernova remnants \citep{Reynolds_1984, Gaensler_2006, Dong_2023}. While some models predict a steep spectral index and luminosity ratio between the radio and X-ray emission of $\gtrsim 8 \times 10^{-3}$ \citep{Murase_2021}, observations of Galactic PWNe suggest otherwise. For example, typically, the radio emission is somewhat flat, with a spectral index of $-0.3 \leq \hat{\beta} \leq 0$. In addition, when observations of radio and X-ray emission are available at the same time the ratio of X-ray luminosity to radio luminosity is $\lesssim 10^{-4}$ \citep{Reynolds_1984}. Unlike PWN emission, we find that the broadband late-time radio emission from SN\,2012ap is steeper, $\hat{\beta} \sim -0.85$, and that the X-ray to radio luminosity ratio is $\sim 1$. In addition, the radii we infer from the SSA analysis, $R \approx 6 \times 10^{17} \, \rm cm$, require high velocities for a newly formed PWN.

Furthermore, the emission from the PWN outflow should illuminate the metal-rich SN ejecta, resulting in broad oxygen emission lines in the optical spectra that is not detected here. This is seen, for example, in SN\,2012au (Wiston et al. in prep) and PWN VT 1137-0377 \citep[see, however, the modeling in \citealt{Omand_2023}]{dan12au, Dong_2023}. The optical spectra we obtained for SN\,2012ap lack broad emission lines. Therefore, we conclude that the late-time radio and X-ray brightening is more likely to be from either CSM density enhancement or an off-axis jet, and not from a PWN outflow.

\subsection{The central-engine context}
\label{subsec: astro_context_discussion}

\begin{figure}[ht]
\centering
\includegraphics[width=\linewidth]{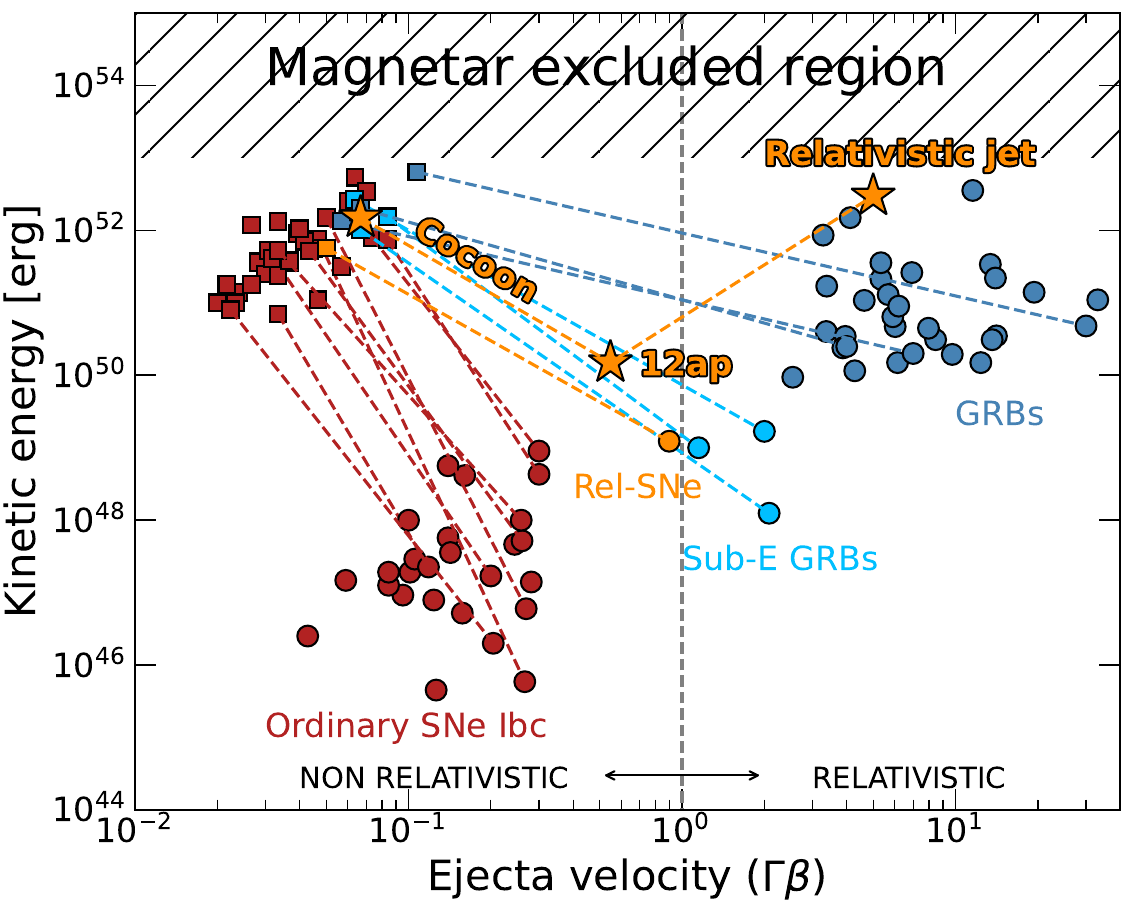}
\caption{Kinetic energy of the ejecta of ordinary type Ibc SNe (red) and Energetic-SNe (E-SNe), a class of explosions that includes GRBs (blue), sub-E GRBs (light blue) and relativistic SNe (orange). This plot is the same as Fig.~2 in \cite{Margutti_2014} but with updated values for SN\,2012ap (orange stars) based on the results of this work. Squares and circles are for the slow-moving and the fast-moving ejecta, respectively. As demonstrated in this plot, the energy of the jet component of SN\,2012ap is similar to other GRBs, suggesting that the main difference between SN\,2012ap and other engine-driven explosions is due to a viewing angle effect.
\label{fig: ek_u_plot}}
\end{figure}

So far, based on the low energy and velocity measured in its early radio emission, and the lack of GRB detection, SN\,2012ap represented an intermediate element of a class between sub-relativistic SNe and fully relativistic GRBs. In addition, the energy-velocity profile at early times is flatter than expected in the purely hydrodynamical scenario and more consistent with a short-lived engine driving the SN explosion \citep{Margutti_2014}. Therefore, it was argued to be one of the weakest observed engine-driven explosion (the other one being SN\,2009bb; \citealt{Soderberg_2010}), where the jet fails to breakout \citep{Margutti_2014}. Our analysis of the new data obtained in the decade since the stellar explosion suggests that a powerful relativistic jet might have been launched from SN\,2012ap, and that the observational peculiarities can be explained with a large viewing angle. In Fig.~\ref{fig: ek_u_plot} we present the phase space of $E_{k}$ and $\Gamma \beta$ (adapted from \citealt{Margutti_2014}). This plot demonstrates that if the late time radio emission is due to a relativistic jet, SN\,2012ap is similar to relativistic GRBs in terms of energy and velocity, and can no longer be considered as an intermediate class.

The implications of our analysis of SN\,2012ap extend beyond its re-classification (if indeed it generated an energetic off-axis jet) and become important in the era of the Einstein Probe. If SN\,2012ap launched a powerful relativistic jet viewed off-axis, then events traditionally identified as engine-driven explosions with jets that fail to break out may instead represent a large population of relativistic SNe whose jets are simply misaligned with our line of sight. The wide-field, high-cadence soft X-ray monitoring enabled by the Einstein Probe is expected to uncover faint, rapidly evolving X-ray transients, with key examples being EP\,240414a \citep{Bright_2025, Srivastav_2025, Zheng_2025} and EP\,250108a \citep{Eyles-Ferris_2025, Srinivasaragavan_2025}. These X-ray transients are typically associated with shock breakout, cocoon emission, or the early interaction of off-axis jets with their environments, and their signals would have been missed by previous GRB-triggered searches \citep{Yuan_2015}.

However, as demonstrated by this study, X-ray detections alone will not allow us to determine the nature of these events. Combining these detections with late-time radio observations, and VLBI observations on different timescales, remain essential for constraining the kinetic energy and geometry of the ejecta, and for distinguishing between spherical mildly-relativistic explosions and relativistic jets observed off-axis. Our results therefore suggest that systematic radio follow-up of Einstein Probe transients will be critical for building an unbiased census of jet production in CCSNe and for establishing whether events like SN\,2012ap represent the low-luminosity tail of the GRB population or a previously hidden majority of off-axis relativistic explosions.

\section{Conclusions}
\label{sec: conc}

In this work we present new radio, optical, and X-ray observations of the 14-year-old relativistic SN\,2012ap and provide detailed analysis of this broadband emission. Our new observations, starting $\sim 9$ years after stellar explosion and spanning over four years, reveal (i) optical spectra consisting of narrow lines, (ii) late-time brightening of the X-ray emission, and (iii) broadband radio rebrightening (the rebrightening in a single radio frequency was first reported by \citealt{Stroh_2021}).

By modeling the radio emission we conclude that the origin of this late-time rebrightening can be explained either by (1) the interaction of the mildly-relativistic SN outflow observed at early times with a density enhancement in the CSM around few$\times 10^{17} \, {\rm cm}$, possibly due to a change in the mass-loss processes during the transition of the progenitor from a RSG to a WR star, or (2) a relativistic narrow jet, with a collimated energy of $E_{k} \lesssim 3 \times 10^{52} \, {\rm erg}$ and an opening angle of $\sim 7.5^{\circ}$, observed off-axis, with a viewing angle of $\theta_{\rm obs} \sim 80^{\circ} - 90^{\circ}$. Our analysis of the X-ray emission points to thermal emission from either the RS or the FS in the CSM. Therefore, the SN ejecta interacting with the CSM naturally explains the overall radio-to-X-ray emission. On the other hand, while the jet+cocoon model naturally explains the radio emission, we still need to invoke either the RS or the FS to explain the X-ray emission. In addition, lateral expansion of the jet is expected but requires faster evolution of the radio emission than the observed evolution. We conclude that direct measurements of the size of the emitting region with our planned VLBA observations of this SN will distinguish between the different physical models presented here.

We emphasize that if the late-time radio emission is due to an off-axis relativistic jet SN\,2012ap can no longer be considered as a ``missing-link" SN. In this context, the only difference between fully relativistic GRBs (represented in blue in Fig.~\ref{fig: ek_u_plot}) and SN\,2012ap is a viewing angle effect, and if we were to observe it on-axis it would have looked similar to other GRBs. This raises the question, are other relativistic SNe (namely SN\,2009bb represented in orange in Fig.~\ref{fig: ek_u_plot}) and sub-energetic GRBs (seen in light blue in Fig.~\ref{fig: ek_u_plot}) similar to regular GRBs but observed off-axis? As seen from our analysis of SN\,2012ap, late-time follow up observations of these events can hold the key to answer this question.

\begin{acknowledgments}
The National Radio Astronomy Observatory (NRAO) is a facility of the National Science Foundation operated under cooperative agreement by Associated Universities, Inc. We thank the NRAO for carrying out the Karl G. Jansky Very Large Array (VLA).
Construction and installation of VLITE was supported by the U.S. Naval Research Laboratory Sustainment Restoration and Maintenance fund.
We thank the staff of the GMRT that made these observations possible. The GMRT is run by the National Centre for Radio Astrophysics of the Tata Institute of Fundamental Research.
This paper makes use of the following ALMA data: ADS/JAO.ALMA\#2021.1.00099.S ALMA is a partnership of ESO (representing its member states), NSF (USA) and NINS (Japan), together with NRC (Canada), NSTC and ASIAA (Taiwan), and KASI (Republic of Korea), in cooperation with the Republic of Chile. The Joint ALMA Observatory is operated by ESO, AUI/NRAO and NAOJ.
This research has made use of the NuSTAR Data Analysis Software (NuSTARDAS) jointly developed by the ASI Space Science Data Center (SSDC, Italy) and the California Institute of Technology (Caltech, USA).
Support for this work was provided by the National Aeronautics and Space Administration through Chandra Award Numbers GO2-23052X and GO2-23039X issued by the Chandra X-ray Center, which is operated by the Smithsonian Astrophysical Observatory for and on behalf of the National Aeronautics Space Administration under contract NAS8-03060.
Basic research in radio astronomy at the U.S. Naval Research Laboratory is supported by 6.1 Base funding.
Some of the data presented herein were obtained at Keck Observatory, which is a private 501(c)3 non-profit organization operated as a scientific partnership among the California Institute of Technology, the University of California, and the National Aeronautics and Space Administration. The Observatory was made possible by the generous financial support of the W. M. Keck Foundation. 
The authors wish to recognize and acknowledge the very significant cultural role and reverence that the summit of Maunakea has always had within the Native Hawaiian community. We are most fortunate to have the opportunity to conduct observations from this mountain. 

R.~M. acknowledges partial support from the National Science Foundation (grant number AST-2224255).
P.~B. acknowledges support by a grant (no. 2024788) from the United States-Israel Binational Science Foundation (BSF), Jerusalem, Israel, by a grant (no. 1649/23) from the Israel Science Foundation and by a NASA grant (80NSSC24K0770).
This research was funded by the National Science Centre, Poland (grant number 2023/49/B/ST9/00066).
FDC acknowledges support from the DGAPA/PAPIIT grant IN113424
D.M. acknowledges support from the National Science Foundation through grants PHY-2209451 and AST-2206532.
MRD acknowledges support from the NSERC through grant RGPIN-2025-06224, the Canada Research Chairs Program, and the Dunlap Institute at the University of Toronto.
RBD acknowledges support from the National Science Foundation under grant 2510568.
W.J.-G.\ is supported by NASA through Hubble Fellowship grant HSTHF2-51558.001-A awarded by the Space Telescope Science Institute, which is operated for NASA by the Association of Universities for Research in Astronomy, Inc., under contract NAS5-26555.
K. D. A. gratefully acknowledges support from the Alfred P. Sloan Foundation.
\end{acknowledgments}

\software{
astropy \citep{2013A&A...558A..33A,2018AJ....156..123A}, 
CASA \citep{CASA},
\texttt{VegasAfterglow} \citep{Wang_2026},
\texttt{emcee} \citep{Foreman_Mackey_2013},
}

\appendix

\section{The position of the cooling spectral break}
\label{subsec: cool_disc}

Our analysis of the late-time radio emission is made under the assumption of $\nu_{\rm c} \simeq 140$ GHz (based on the observation with ALMA) which implies a deviation from equipartition ($\epsilon_{e}/\epsilon_{B} \sim 0.1$ in our analysis). However, the bottom panel of Fig.~\ref{fig: SN2012ap_radio} reveals that the X-ray emission matches the tail of the low-GHz emission (up to $\sim 10 \, {\rm GHz}$). This can suggest that the cooling frequency is actually at higher frequencies than we observe with ALMA. If the cooling frequency is around the X-ray band, the deviation from equipartition will increase significantly. The synchrotron cooling frequency is
\begin{align}
\label{eq: cool_freq}
    \nu_{\rm c} = \frac{8 \pi m_{\rm e} c q_{\rm e}}{\sigma_{\rm T}^2 \Gamma B^3 t_{\rm cool}^2}
\end{align}
where $m_{\rm e}$ and $q_{\rm e}$ are the mass and charge of the electron, respectively, $\sigma_{\rm T}$ is Thomson cross-section, $B$ is the magnetic field strength, and $t_{\rm cool}$ is the cooling time of the electrons. In the non-relativistic regime discussed in \S~\ref{subsub:radioFS}, the magnetic field strength is fixed for a set of fixed $F_{\rm \nu_{\rm sa}}$, $\nu_{\rm sa}$, and $p$, and varies for different energy partition ratios $\alpha \equiv \epsilon_{\rm e} / \epsilon_{\rm B}$ (see Eq.~\ref{eq: bfield_chevalier}; note that we assume $\Gamma \sim 1$). Therefore, varying $\alpha$ and/or $t_{\rm cool}$ is the only way to achieve a cooling break frequency at, or above, the X-ray band. 

\begin{figure}[ht]
\centering
\includegraphics[width=\linewidth]{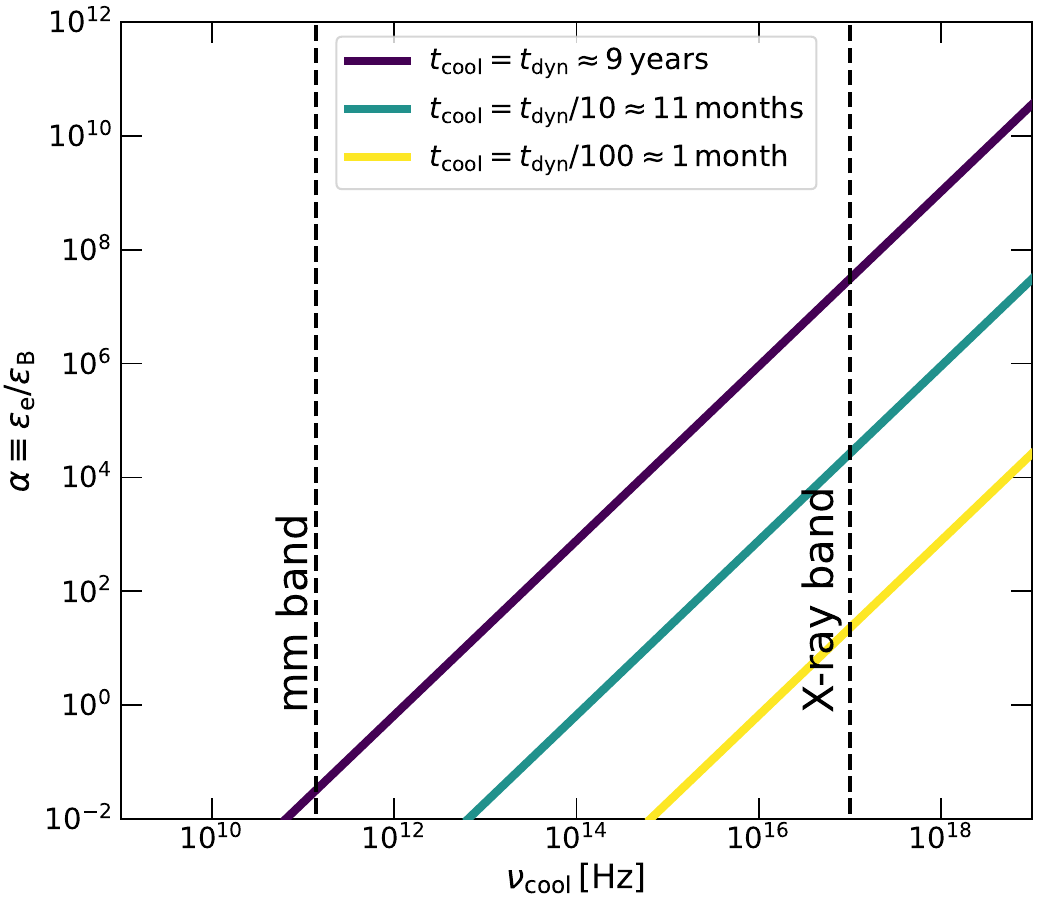}
\caption{The ratio of the energy in the relativistic electrons, $\epsilon_{\rm e}$, over the energy in the magnetic fields, $\epsilon_{\rm B}$, vs. the position of the cooling break frequency, $\nu_{\rm c}$, for different cooling times, $t_{\rm cool}$. For reference, we plot in dashed vertical lines the position of the cooling break inferred in the mm-bands, and the position of the X-ray band. This analysis implies that extreme $\epsilon_{\rm e} / \epsilon_{\rm B}$ ratios are needed in order to reconcile the position of the cooling frequency with the naive extrapolation of the radio SEDs to the X-ray band (as seen in Fig.~\ref{fig: SN2012ap_radio}).
\label{fig: cool_freq}}
\end{figure}

So far we assumed that the cooling time is the dynamical time, $t_{\rm dyn}$. This assumption implies that the electrons are cooled on timescales similar to the time since explosion. In Fig.~\ref{fig: cool_freq} we present the ratio of energy partition, $\alpha$, vs. the position of the cooling break frequency, $\nu_{\rm c}$, for different cooling timescales, $t_{\rm cool}$. This analysis indicates nonphysical ratios for the long cooling timescales, $\alpha \gtrsim 10^{7}$ and $10^{4}$ for $t_{\rm cool} \approx 9$ and $1$ years, respectively, in order to get a cooling break above the X-ray band. Alternatively, more reasonable ratios of $\alpha < 10^{3}$ are allowing for $\nu_{\rm c} \gtrsim 10^{17}$ Hz if we assume $t_{\rm cool} \sim$\,few months. However, this means that the vast majority of the electrons are not emitting in synchrotron. Therefore, we conclude that while the X-ray emission seems to be a natural extension of the synchrotron emission, the inferred physical parameters (especially the energy partition ratio) are extreme, and it is more likely that the emission in X-rays arises from a different mechanism (as discussed in detail in \S~\ref{subsub:XraysRS}).

\section{Data tables}
\label{sec: tables}

In this section we attach the tables containing data and best-fitting parameters.

\begin{deluxetable*}{ccccccccc}
\tablecaption{CXO and NuSTAR X-ray observations of SN\,2012ap and inferred spectral parameters.}
\tablehead{
\colhead{Obs ID} & \colhead{$\delta t$} & \colhead{Count-Rate$^a$} & \colhead{Significance$^b$} & \colhead{Photon Index}&
\colhead{Log Unabsorbed Flux$^c$} &
\colhead{Spacecraft}  \\
\colhead{} & \colhead{$\rm \left[ d \right]$} &\colhead{$10^{-4}\,[\rm{c\,s^{-1}}$]} & \colhead{$\sigma$} & &\colhead{$[\rm{erg\,s^{-1}\,cm^{-2}}$]} }
\startdata 
\hline
13785&24 & $<8$ & -- & 2$^d$ & $<-14.1$ & CXO\\
25181& 3596 & $6.0\pm0.2$  &5.0 & $1.89^{+1.5}_{-0.4}$&$-13.9^{+0.4}_{-0.1}$ & CXO\\
25216& 3999& $5.6 \pm 2.0$ &  $3.9$ &  \rdelim\}{2}{*}[\,$2.2^{+0.9}_{-0.6}$ ]  & \rdelim\}{2}{*}[\,$-13.7^{+0.3}_{-0.1}$ ] & CXO\\ [3pt]
25217& 4082& $8.4\pm2.4$ & 6.3 &\\
80802504002 &3998& $<2.8$  &-- &2$^d$ & $<-13.8$&NuSTAR \\
80802504004 &4081& $<2.6$  &-- &2$^d$ & $<-13.8$&NuSTAR \\
\hline
\enddata
\tablecomments{$^a$ 3$\sigma$ limit; Background-subtracted; 0.5--8 keV for CXO, 8-20 keV for NuSTAR, both modules. \\ $^b$  0.5--8 keV. \\ $^c$  CXO: 0.3--10 keV and  $\rm{NH_{\rm{int}}}=0\,\rm{cm^{-2}}$. NuSTAR: 8--20 keV. \\$^{d}$ Assumed.
\label{Tab:Xraydata}}
\end{deluxetable*}

\begin{deluxetable*}{cccccc}
\tablecaption{Summary of the radio flux measurements.}
\tablehead{
\colhead{Observation date} & \colhead{$\Delta t$} & \colhead{$\nu$} & \colhead{$F_{\rm \nu}$} &
\colhead{Telescope (project code)} \\
\colhead{[YYYY-MM-DD]} & \colhead{$\rm \left[ Days \right]$} & \colhead{$\rm \left[ GHz \right]$} & \colhead{$\rm \left[ mJy \right]$} & \colhead{}}
\startdata
2017-09-30 & $2064$ & $3.0$ & $3.9 \pm 0.4$ & VLA (VLASS) \\
2020-08-16 & $3115$ & $3.0$ & $4.2 \pm 0.5$ & VLA (VLASS) \\
2021-01-30 & $3282$ & $0.341$ & $7.7 \pm 1.5$ & VLA (VLITE) \\
2021-01-30 & $3282$ & $1.26$ & $8.7 \pm 0.5$ & VLA:A (20B-279) \\
2021-01-30 & $3282$ & $1.78$ & $6.9 \pm 0.4$ & VLA:A (20B-279) \\
2021-01-30 & $3282$ & $2.50$ & $5.3 \pm 0.3$ & VLA:A (20B-279) \\
2021-01-30 & $3282$ & $3.67$ & $4.1 \pm 0.2$ & VLA:A (20B-279) \\
2021-01-30 & $3282$ & $4.78$ & $3.1 \pm 0.2$ & VLA:A (20B-279) \\
2021-01-30 & $3282$ & $5.76$ & $2.7 \pm 0.1$ & VLA:A (20B-279) \\
2021-01-30 & $3282$ & $6.69$ & $2.3 \pm 0.1$ & VLA:A (20B-279) \\
2021-01-30 & $3282$ & $7.71$ & $2.0 \pm 0.1$ & VLA:A (20B-279) \\
2022-06-09 & $3624$ & $211$ & $0.086 \pm 0.009$ & ALMA (2021.1.00099.S) \\
2022-08-15 & $3844$ & $0.33$ & $10.4 \pm 1.2$ & GMRT \\
2022-08-15 & $3844$ & $0.65$ & $13.6 \pm 1.4$ & GMRT \\
2022-08-15 & $3844$ & $1.37$ & $7.5 \pm 0.8$ & GMRT \\
2023-02-26 & $4039$ & $3.0$ & $4.2 \pm 0.5$ & VLA:A (VLASS) \\
2024-10-30 & $4651$ & $0.341$ & $10.4 \pm 1.9$ & VLA:A (VLITE) \\
2024-10-30 & $4651$ & $1.5$ & $7.9 \pm 0.8$ & VLA:A (24B-311) \\
2024-10-30 & $4651$ & $3.0$ & $3.6 \pm 0.4$ & VLA:A (24B-311) \\
2024-10-30 & $4651$ & $6.0$ & $2.4 \pm 0.3$ & VLA:A (24B-311) \\
2024-10-30 & $4651$ & $10.0$ & $1.5 \pm 0.2$ & VLA:A (24B-311) \\
\enddata
\tablecomments{Summary of the late-time radio flux density measurements for SN\,2012ap. $\delta t$ is the time since explosion, $\nu$ is the observed central frequency, $F_{\rm \nu}$ is the flux density (and $1 \sigma$ uncertainty). In the ``Telescope'' column we specify the array configuration after the colon, and the project (in parentheses).
\label{tab: radio_observations}}
\end{deluxetable*}

\begin{deluxetable}{cccc}
\tablecaption{Broken power-law fits to individual SEDs.}
\tablehead{
\colhead{$\delta t$} & \colhead{$F_{\nu_{\rm sa}}$}& \colhead{$\nu_{\rm sa}$} & \colhead{$p$} \\
\colhead{$\rm \left[ Days \right]$} & \colhead{$\rm \left[ mJy \right]$} & \colhead{$\rm \left[ GHz \right]$}& \colhead{}}
\startdata 
$12$ & $9.9 \pm 0.8$ & $10.2 \pm 0.6$ & $3.15^{\dagger}$ \\
$18$ & $9.1 \pm 0.7$ & $7.85 \pm 0.50$ & $3.15 \pm 0.30$ \\
$27$ & $7.4 \pm 0.4$ & $5.35 \pm 0.20$ & $3.15^{\dagger}$ \\
$38$ & $6.5 \pm 0.3$ & $3.9 \pm 0.2$ & $3.15^{\dagger}$ \\
\hline
$3282$ & $20 \pm 1$ & $0.47 \pm 0.03$ & $2.62 \pm 0.06$ \\
$3844-4039$ & $22 \pm 2$ & $0.39 \pm 0.04$ & $2.7 \pm 0.2$ \\
$4651$ & $22 \pm 3$ & $0.42 \pm 0.05$ & $2.7 \pm 0.2$ \\
\enddata
\tablecomments{Best-fitting parameters of the radio SEDs using Eq.~\ref{eq: bpl_function}. $\dagger$ marks observations for which the power-law index of the electron distribution is fixed (see discussion in \S\ref{subsec: density_enhancement}). The reported times and peak frequencies are in the frame of the observer.
\label{tab: bpl_fits_parameters}}
\end{deluxetable}

\begin{deluxetable*}{ccccccccc}
\tablecaption{Physical parameters from a non-relativistic SSA analysis of individual SED fits.}
\tablehead{
\colhead{} & \colhead{$\delta t$} & \colhead{$R$} & \colhead{$B$} &
\colhead{$\Gamma\beta$} & \colhead{$\rho_{\rm csm}$} & \colhead{$U_{\rm ps}$} & \colhead{$\nu_{\rm m}$} & \colhead{$\nu_{\rm c}$} \\
\colhead{} & \colhead{$\rm \left[ Days \right]$} & \colhead{$\rm \left[ 10^{16} \, cm \right]$} & \colhead{$\rm \left[ G \right]$} & \colhead{} & \colhead{$\rm \left[ 10^{-22} \,g\, cm^{-3} \right]$} & \colhead{$\rm \left[ 10^{48} \,erg \right]$} & \colhead{$\rm \left[ GHz \right]$} & \colhead{$\rm \left[ GHz \right]$}}
\startdata 
\hline
$\epsilon_{\rm e} = \epsilon_{\rm B} = 0.1$ & $12$ & $1.7 \pm 0.2$ & $0.93^{+0.18} _{-0.14}$ & $0.66^{+0.12} _{-0.09}$ & $12^{+3}_{-2}$ & $3.9^{+1.6} _{-1.1}$ & $1.2 \pm 0.3$ & $1900^{+1100} _{-800}$ \\
& $18$ & $2.2^{+0.3} _{-0.2}$ & $0.7 \pm 0.1$ & $0.55^{+0.09} _{-0.07}$ & $9 \pm 2$ & $5.2^{+2.1} _{-1.4}$ & $0.43^{+0.10} _{-0.07}$ & $1800^{+1100} _{-700}$ \\
& $27$ & $3.0 \pm 0.3$ & $0.51^{+0.09} _{-0.07}$ & $0.47^{+0.07} _{-0.06}$ & $5.6^{+1.1} _{-0.8}$ & $5.8^{+2.3} _{-1.5}$ & $0.16 \pm 0.03$ & $2400^{+1400} _{-900}$ \\
& $38$ & $3.8 \pm 0.4$ & $0.37^{+0.07} _{-0.05}$ & $0.42^{+0.06} _{-0.05}$ & $3.6^{+0.8} _{-0.7}$ & $6.9^{+2.7} _{-1.9}$ & $0.08^{+0.02} _{-0.01}$ & $3000^{+1800} _{-1200}$ \\
& $3282$ & $45 \pm 3$ & $0.030 \pm 0.002$ & $0.053 \pm 0.004$ & $1.3^{+0.4} _{-0.3}$ & $75^{+10} _{-9}$ & $< 10^{-2}$ & $700^{+140} _{-120}$ \\
& $3844-4039$ & $55^{+5} _{-4}$ & $0.025 \pm 0.003$ & $0.055^{+0.006} _{-0.004}$ & $0.8 \pm 0.3$ & $94 \pm 18$ & $< 10^{-2}$ & $940^{+370} _{-220}$ \\
& $4651$ & $53^{+6} _{-5}$ & $0.027 \pm 0.003$ & $0.044^{+0.005} _{-0.004}$ & $1.5^{+0.7} _{-0.5}$ & $110 \pm 20$ & $< 10^{-2}$ & $520^{+230} _{-140}$ \\
\hline
$\epsilon_{\rm e} = 10^{-3}, \, \epsilon_{\rm B} = 10^{-2}$ & $12$ & $1.9 \pm 0.2$ & $1.5^{+0.3} _{-0.2}$ & $0.80^{+0.15} _{-0.10}$ & $240^{+50}_{-40}$ & $150^{+50} _{-40}$ & $<10^{-2}$ & $450^{+250} _{-170}$ \\
& $18$ & $2.5 \pm 0.3$ & $1.2 \pm 0.2$ & $0.64^{+0.11} _{-0.09}$ & $190^{+50} _{-40}$ & $190^{+70} _{-50}$ & $<10^{-2}$ & $430^{+240} _{-160}$ \\
& $27$ & $3.3^{+0.4} _{-0.3}$ & $0.8 \pm 0.1$ & $0.54^{+0.08} _{-0.07}$ & $120 \pm 20$ & $220^{+80} _{-50}$ & $<10^{-2}$ & $570^{+300} _{-200}$ \\
& $38$ & $4.3^{+0.5} _{-0.4}$ & $0.61^{+0.10} _{-0.08}$ & $0.49^{+0.07} _{-0.06}$ & $75^{+15} _{-10}$ & $260^{+90} _{-60}$ & $<10^{-2}$ & $710^{+380} _{-260}$ \\
& $3282$ & $51 \pm 4$ & $0.051 \pm 0.003$ & $0.061 \pm 0.004$ & $28^{+8} _{-6}$ & $3000 \pm 40$ & $< 10^{-2}$ & $160 \pm 30$ \\
& $3844-4039$ & $62^{+6} _{-5}$ & $0.042 \pm 0.004$ & $0.063^{+0.006} _{-0.005}$ & $17^{+7} _{-6}$ & $3800 \pm 800$ & $< 10^{-2}$ & $200^{+80} _{-50}$ \\
& $4651$ & $61^{+7} _{-6}$ & $0.045 \pm 0.005$ & $0.050^{+0.006} _{-0.005}$ & $31^{+15} _{-10}$ & $4200 \pm 900$ & $< 10^{-3}$ & $120^{+50} _{-30}$ \\
\hline
\enddata
\tablecomments{The physical parameters of the outflows and their environment based on the broken power-law fits to the individual radio SEDs (\S\ref{subsec: density_enhancement}). The radius and magnetic field strength are calculated using Eq. \ref{eq: radius_chevalier} and \ref{eq: bfield_chevalier}, respectively. The proper velocity, $\Gamma\beta$, is defined by $R/ \delta t \equiv \beta c$, and $\Gamma \equiv \frac{1}{1-\beta^2}$. The CSM density is calculated using Eq. \ref{eq: density_bfield}. The post-shock energy and it is calculated assuming $U_{\rm ps} = \frac{1}{\epsilon_{\rm B}}\frac{B^2}{8 \pi} V$, where $V = f \frac{4 \pi}{3} R^3$ is the volume of the emitting region. These reported values assume $\epsilon_{\rm e} = \epsilon_{\rm B} = 0.1$ for $\delta t = 12-38$\,d, and $\epsilon_{\rm e} = 0.01$ and $\epsilon_{\rm B} = 0.1$ for $\delta t = 3282-4651$\,d (see details in \S\ref{subsec: density_enhancement}).
\label{tab: equipartition_parameters}}
\end{deluxetable*}

\begin{deluxetable*}{ccccc}[ht]
\tablecaption{A summary of the off-axis Jet+cocoon model parameters.}
\tablehead{
\colhead{Parameter} & \colhead{Units} & \colhead{Definition} &
\colhead{Priors} &
\colhead{Best-fitting value}}
\startdata
Jet component & & non-spreading jet & & \\
\hline
$\log_{10} \left( E_{k, \, True, \, j}\right)$ & ${\rm \log_{10} \left( erg \right)}$ & \pbox{2.75in}{Collimated corrected energy of the jet, log scale.} & $\left[ 48-\log_{\rm 10} \left( 3 \times 10^{52}\right)\right]$ & $52-\log_{10} \left( 3 \times 10^{52} \right)$ \\
$\theta_{0, \, j}$ & (rad) & \pbox{2.75in}{Opening angle of the jet.} & $\left[ 0.017 - 1.05\right]$ & $0.13^{+0.04} _{-0.05}$ \\
$\Gamma_{0, \, j}$ & - & \pbox{2.75in}{Initial Lorentz factor of the jet.} & $\left[ 2-30\right]$ & $\geq 5$ \\
$\log_{\rm 10} \left(\epsilon_{e, \, j}\right)$ & - & \pbox{2.75in}{Fraction of energy in the relativistic electrons, log scale.} & $\left[ -5 - \left( -0.5 \right) \right]$ & $-2.8^{+0.3} _{-0.2}$ \\
$\log_{\rm 10} \left(\epsilon_{B, \, j}\right)$ & - & \pbox{2.75in}{Fraction of energy in the magnetic fields, log scale.} & $\left[ -5 - \left( -0.5 \right) \right]$ & $-1.0^{+0.2} _{-0.4}$ \\
$p_{j}$ & - & \pbox{2.75in}{Power-law index of the relativistic electrons} & $\left[ 2.01-4 \right]$ & $2.70 \pm 0.03$ \\ \\
\hline
Cocoon component & & non-spreading jet & & \\
\hline 
$\log_{10} \left( E_{k, \, iso, \, c}\right)$ & ${\rm \log_{10} \left( erg \right)}$ & \pbox{2.75in}{Isotropic equivalent energy of the cocoon, log scale.} & $\left[ 48-52\right]$ & $50.2 \pm 0.1$ \\
$\theta_{0, \, c}$ & (degree) & \pbox{2.75in}{Opening angle of the cocoon.} & fixed & $60^{\circ}$ \\
$\Gamma_{0, \, c}$ & - & \pbox{2.75in}{Initial Lorentz factor of the cocoon.} & $\left[ 1-2 \right]$ & $1.14 \pm 0.02$ \\
$\epsilon_{e, \, c}$ & - & \pbox{2.75in}{Fraction of energy in the relativistic electrons.} & fixed & $0.1$ \\
$\epsilon_{B, \, c}$ & - & \pbox{2.75in}{Fraction of energy in the magnetic fields.} & fixed & $0.1$ \\
$p_{\rm c}$ & - & \pbox{2.75in}{Power-law index of the relativistic electrons} & $\left[ 2.01-4 \right]$ & $2.35^{+0.15} _{-0.12}$ \\ \\
\hline
Environment & & $\rho = A_{\rm *} r^{-k} + n_{\rm ISM} \, m_{\rm p}$ & & \\
\hline
$A_{\rm *}$ & $\left( 5 \times 10^{11} \, g \, cm^{-3} \right)$ & \pbox{2.75in}{Normalization factor of the CSM density.} & $\left[ 0.01-100 \right]$ & $2.0^{+0.6} _{-0.5}$ \\
$k$ & - & \pbox{2.75in}{Power-law index of the CSM density profile.} & $\left[ 0-2.9 \right]$ & $1.55 \pm 0.12$ \\
$n_{\rm ISM}$ & $\left( cm^{-3} \right)$ & \pbox{2.75in}{Number density of the ISM.} & fixed & $1$ \\ \\
\hline
Observer & & & & \\
\hline
$\theta_{\rm obs}$ & (rad) & \pbox{2.75in}{Viewing angle.} & $\left[ 1.05-1.57 \right]$ & $1.38-1.57$ \\
\enddata
\tablecomments{The physical parameters of the off-axis jet+cocoon \texttt{VegasAfterglow} model \citep{Wang_2026} we use to fit both the early and late-time radio data. We use flat priors, $200$ walkers, and $10000$ steps with a burn-in of $1000$ in our implementation of MCMC. \texttt{VegasAfterglow} uses the isotropic equivalent energy, $E_{k, \, iso}$, however, for the jet component we constrain the collimated corrected energy to be up to $E_{k, \, True} \leq 3 \times 10^{52} \, \rm erg$ as higher values are unlikely. To do so we assume in our fitting code that $E_{k, \, True, \, j} \simeq 0.5 \, \theta_{0, \, j}^2 \, E_{k, \, iso, \, j}$. See detailed discussion in \S\ref{subsec: jet_cocoon}.
\label{tab: jet_cocoon_model_parameters}}
\end{deluxetable*}

\section{Corner plots}
\label{sec: corner_plots}

In this section we present the corner plot of the MCMC analysis presented in \S\ref{subsec: jet_cocoon}.

\begin{figure*}
\centering
\includegraphics[width=\linewidth]{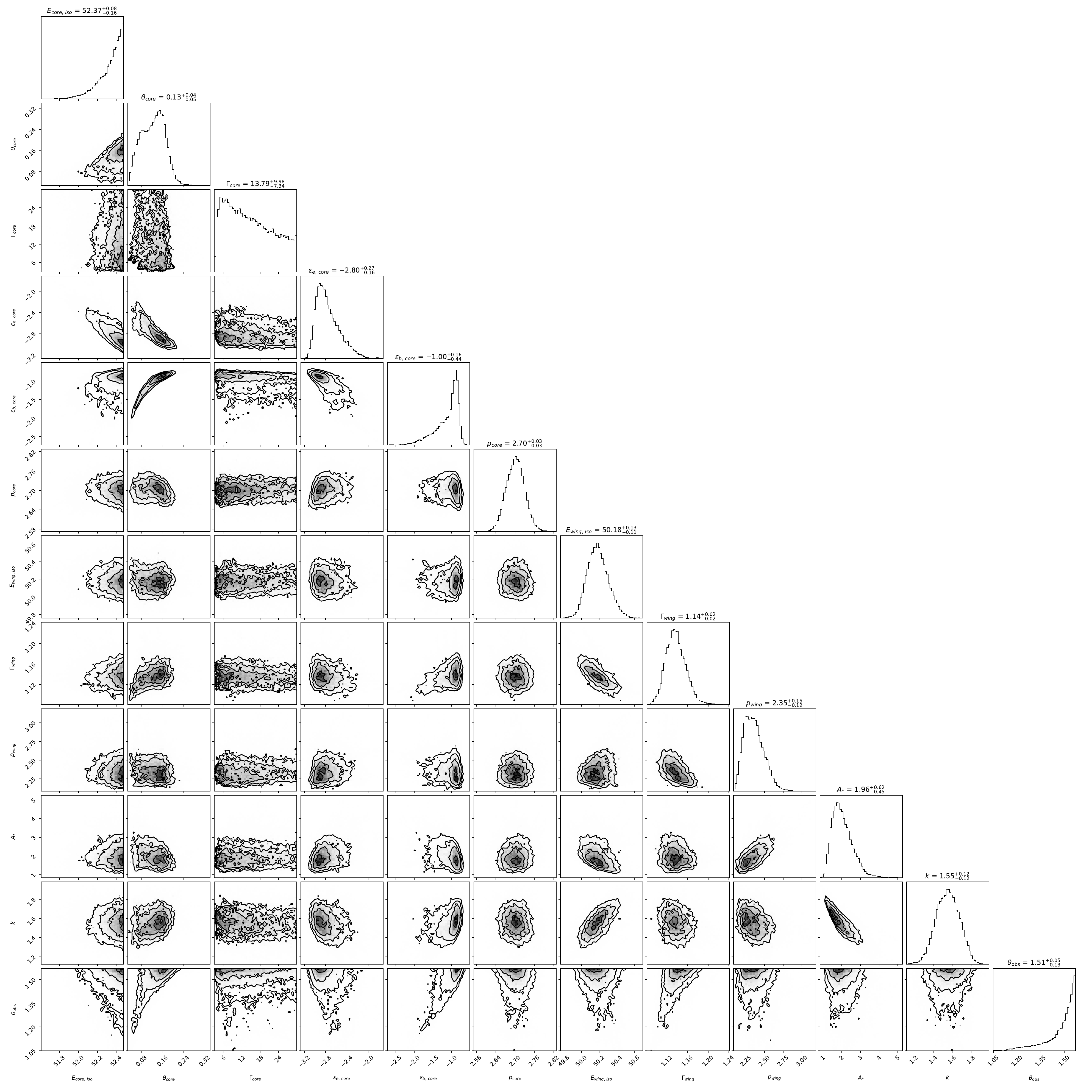}
\caption{Corner plots of the posterior distributions for the jet+cocoon model using \texttt{VegasAfterglow} code. As seen from these plots, the collimated corrected energy of the jet is reaching the highest value we impose as an upper limit for our prior, $3 \times 10^{52} \, \rm erg$, the initial Lorentz factor of the jet is unconstrained and therefore serves as a lower limit. In addition, the observing angle is also reaching the highest value of the prior, $\theta_{\rm obs} = 90^{\circ}$. In Fig.~\ref{fig: jet_cocoon_fit} we present the radio SEDs with lines drawns from these posterior distributions.
\label{fig: jet_cocoon_corner}}
\end{figure*}

\bibliography{sample701}{}

\begin{thebibliography}{}
\expandafter\ifx\csname natexlab\endcsname\relax\def\natexlab#1{#1}\fi
\providecommand{\url}[1]{\href{#1}{#1}}
\providecommand{\dodoi}[1]{doi:~\href{http://doi.org/#1}{\nolinkurl{#1}}}
\providecommand{\doeprint}[1]{\href{http://ascl.net/#1}{\nolinkurl{http://ascl.net/#1}}}
\providecommand{\doarXiv}[1]{\href{https://arxiv.org/abs/#1}{\nolinkurl{https://arxiv.org/abs/#1}}}

\bibitem[{G.~E. {Anderson} {et~al.}(2017){Anderson}, {Horesh}, {Mooley}, {Rushton}, {Fender}, {Staley}, {Argo}, {Beswick}, {Hancock}, {P{\'e}rez-Torres}, {Perrott}, {Plotkin}, {Pretorius}, {Rumsey}, \& {Titterington}}]{Anderson_2017}
{Anderson}, G.~E., {Horesh}, A., {Mooley}, K.~P., {et~al.} 2017, \bibinfo{title}{{The peculiar mass-loss history of SN 2014C as revealed through AMI radio observations},} \mnras, 466, 3648, \dodoi{10.1093/mnras/stw3310}

\bibitem[{ {Astropy Collaboration} {et~al.}(2013){Astropy Collaboration}, {Robitaille}, {Tollerud}, {Greenfield}, {Droettboom}, {Bray}, {Aldcroft}, {Davis}, {Ginsburg}, {Price-Whelan}, {Kerzendorf}, {Conley}, {Crighton}, {Barbary}, {Muna}, {Ferguson}, {Grollier}, {Parikh}, {Nair}, {Unther}, {Deil}, {Woillez}, {Conseil}, {Kramer}, {Turner}, {Singer}, {Fox}, {Weaver}, {Zabalza}, {Edwards}, {Azalee Bostroem}, {Burke}, {Casey}, {Crawford}, {Dencheva}, {Ely}, {Jenness}, {Labrie}, {Lim}, {Pierfederici}, {Pontzen}, {Ptak}, {Refsdal}, {Servillat}, \& {Streicher}}]{2013A&A...558A..33A}
{Astropy Collaboration}, {Robitaille}, T.~P., {Tollerud}, E.~J., {et~al.} 2013, \bibinfo{title}{{Astropy: A community Python package for astronomy},} \aap, 558, A33, \dodoi{10.1051/0004-6361/201322068}

\bibitem[{ {Astropy Collaboration} {et~al.}(2018){Astropy Collaboration}, {Price-Whelan}, {Sip{\H{o}}cz}, {G{\"u}nther}, {Lim}, {Crawford}, {Conseil}, {Shupe}, {Craig}, {Dencheva}, {Ginsburg}, {VanderPlas}, {Bradley}, {P{\'e}rez-Su{\'a}rez}, {de Val-Borro}, {Aldcroft}, {Cruz}, {Robitaille}, {Tollerud}, {Ardelean}, {Babej}, {Bach}, {Bachetti}, {Bakanov}, {Bamford}, {Barentsen}, {Barmby}, {Baumbach}, {Berry}, {Biscani}, {Boquien}, {Bostroem}, {Bouma}, {Brammer}, {Bray}, {Breytenbach}, {Buddelmeijer}, {Burke}, {Calderone}, {Cano Rodr{\'\i}guez}, {Cara}, {Cardoso}, {Cheedella}, {Copin}, {Corrales}, {Crichton}, {D'Avella}, {Deil}, {Depagne}, {Dietrich}, {Donath}, {Droettboom}, {Earl}, {Erben}, {Fabbro}, {Ferreira}, {Finethy}, {Fox}, {Garrison}, {Gibbons}, {Goldstein}, {Gommers}, {Greco}, {Greenfield}, {Groener}, {Grollier}, {Hagen}, {Hirst}, {Homeier}, {Horton}, {Hosseinzadeh}, {Hu}, {Hunkeler}, {Ivezi{\'c}}, {Jain}, {Jenness}, {Kanarek}, {Kendrew}, {Kern}, {Kerzendorf}, {Khvalko}, {King}, {Kirkby}, {Kulkarni},
  {Kumar}, {Lee}, {Lenz}, {Littlefair}, {Ma}, {Macleod}, {Mastropietro}, {McCully}, {Montagnac}, {Morris}, {Mueller}, {Mumford}, {Muna}, {Murphy}, {Nelson}, {Nguyen}, {Ninan}, {N{\"o}the}, {Ogaz}, {Oh}, {Parejko}, {Parley}, {Pascual}, {Patil}, {Patil}, {Plunkett}, {Prochaska}, {Rastogi}, {Reddy Janga}, {Sabater}, {Sakurikar}, {Seifert}, {Sherbert}, {Sherwood-Taylor}, {Shih}, {Sick}, {Silbiger}, {Singanamalla}, {Singer}, {Sladen}, {Sooley}, {Sornarajah}, {Streicher}, {Teuben}, {Thomas}, {Tremblay}, {Turner}, {Terr{\'o}n}, {van Kerkwijk}, {de la Vega}, {Watkins}, {Weaver}, {Whitmore}, {Woillez}, {Zabalza}, \& {Astropy Contributors}}]{2018AJ....156..123A}
{Astropy Collaboration}, {Price-Whelan}, A.~M., {Sip{\H{o}}cz}, B.~M., {et~al.} 2018, \bibinfo{title}{{The Astropy Project: Building an Open-science Project and Status of the v2.0 Core Package},} \aj, 156, 123, \dodoi{10.3847/1538-3881/aabc4f}

\bibitem[{E. {Berger} {et~al.}(2003){Berger}, {Kulkarni}, {Frail}, \& {Soderberg}}]{Berger_2003}
{Berger}, E., {Kulkarni}, S.~R., {Frail}, D.~A., \& {Soderberg}, A.~M. 2003, \bibinfo{title}{{A Radio Survey of Type Ib and Ic Supernovae: Searching for Engine-driven Supernovae},} \apj, 599, 408, \dodoi{10.1086/379214}

\bibitem[{M.~F. {Bietenholz} {et~al.}(2021){Bietenholz}, {Bartel}, {Argo}, {Dua}, {Ryder}, \& {Soderberg}}]{Bietenholz_2021}
{Bietenholz}, M.~F., {Bartel}, N., {Argo}, M., {et~al.} 2021, \bibinfo{title}{{The Radio Luminosity-risetime Function of Core-collapse Supernovae},} \apj, 908, 75, \dodoi{10.3847/1538-4357/abccd9}

\bibitem[{M.~F. {Bietenholz} {et~al.}(2018){Bietenholz}, {Kamble}, {Margutti}, {Milisavljevic}, \& {Soderberg}}]{Bietenholz_2018}
{Bietenholz}, M.~F., {Kamble}, A., {Margutti}, R., {Milisavljevic}, D., \& {Soderberg}, A. 2018, \bibinfo{title}{{SN 2014C: VLBI images of a supernova interacting with a circumstellar shell},} \mnras, 475, 1756, \dodoi{10.1093/mnras/stx3194}

\bibitem[{G. {Birenbaum} {et~al.}(2024){Birenbaum}, {Gill}, {Bromberg}, {Beniamini}, \& {Granot}}]{Birenbaum_2024}
{Birenbaum}, G., {Gill}, R., {Bromberg}, O., {Beniamini}, P., \& {Granot}, J. 2024, \bibinfo{title}{{Afterglow Linear Polarization Signatures from Shallow GRB Jets: Implications for Energetic GRBs},} \apj, 974, 308, \dodoi{10.3847/1538-4357/ad7393}

\bibitem[{G. {Birenbaum} {et~al.}(2026){Birenbaum}, {Granot}, \& {Beniamini}}]{Birenbaum_2026}
{Birenbaum}, G., {Granot}, J., \& {Beniamini}, P. 2026, \bibinfo{title}{{Afterglow linear polarization signatures from steep GRB jets: Implications for orphan afterglows},} \aap, 706, A145, \dodoi{10.1051/0004-6361/202556805}

\bibitem[{D. {Brethauer} {et~al.}(2022){Brethauer}, {Margutti}, {Milisavljevic}, {Bietenholz}, {Chornock}, {Coppejans}, {De Colle}, {Hajela}, {Terreran}, {Vargas}, {DeMarchi}, {Harris}, {Jacobson-Gal{\'a}n}, {Kamble}, {Patnaude}, \& {Stroh}}]{Brethauer22}
{Brethauer}, D., {Margutti}, R., {Milisavljevic}, D., {et~al.} 2022, \bibinfo{title}{{Seven Years of Coordinated Chandra-NuSTAR Observations of SN 2014C Unfold the Extreme Mass-loss History of Its Stellar Progenitor},} \apj, 939, 105, \dodoi{10.3847/1538-4357/ac8b14}

\bibitem[{J.~S. {Bright} {et~al.}(2025){Bright}, {Carotenuto}, {Fender}, {Choza}, {Mummery}, {Jonker}, {Smartt}, {DeBoer}, {Farah}, {Matthews}, {Pollak}, {Rhodes}, \& {Siemion}}]{Bright_2025}
{Bright}, J.~S., {Carotenuto}, F., {Fender}, R., {et~al.} 2025, \bibinfo{title}{{The Radio Counterpart to the Fast X-Ray Transient EP240414a},} \apj, 981, 48, \dodoi{10.3847/1538-4357/adaaef}

\bibitem[{ {CASA Team} {et~al.}(2022){CASA Team}, {Bean}, {Bhatnagar}, {Castro}, {Donovan Meyer}, {Emonts}, {Garcia}, {Garwood}, {Golap}, {Gonzalez Villalba}, {Harris}, {Hayashi}, {Hoskins}, {Hsieh}, {Jagannathan}, {Kawasaki}, {Keimpema}, {Kettenis}, {Lopez}, {Marvil}, {Masters}, {McNichols}, {Mehringer}, {Miel}, {Moellenbrock}, {Montesino}, {Nakazato}, {Ott}, {Petry}, {Pokorny}, {Raba}, {Rau}, {Schiebel}, {Schweighart}, {Sekhar}, {Shimada}, {Small}, {Steeb}, {Sugimoto}, {Suoranta}, {Tsutsumi}, {van Bemmel}, {Verkouter}, {Wells}, {Xiong}, {Szomoru}, {Griffith}, {Glendenning}, \& {Kern}}]{CASA}
{CASA Team}, {Bean}, B., {Bhatnagar}, S., {et~al.} 2022, \bibinfo{title}{{CASA, the Common Astronomy Software Applications for Radio Astronomy},} \pasp, 134, 114501, \dodoi{10.1088/1538-3873/ac9642}

\bibitem[{S. {Chakraborti} {et~al.}(2015){Chakraborti}, {Soderberg}, {Chomiuk}, {Kamble}, {Yadav}, {Ray}, {Hurley}, {Margutti}, {Milisavljevic}, {Bietenholz}, {Brunthaler}, {Pignata}, {Pian}, {Mazzali}, {Fransson}, {Bartel}, {Hamuy}, {Levesque}, {MacFadyen}, {Dittmann}, {Krauss}, {Briggs}, {Connaughton}, {Yamaoka}, {Takahashi}, {Ohno}, {Fukazawa}, {Tashiro}, {Terada}, {Murakami}, {Goldsten}, {Barthelmy}, {Gehrels}, {Cummings}, {Krimm}, {Palmer}, {Golenetskii}, {Aptekar}, {Frederiks}, {Svinkin}, {Cline}, {Mitrofanov}, {Golovin}, {Litvak}, {Sanin}, {Boynton}, {Fellows}, {Harshman}, {Enos}, {von Kienlin}, {Rau}, {Zhang}, \& {Savchenko}}]{Chakraborti_2015}
{Chakraborti}, S., {Soderberg}, A., {Chomiuk}, L., {et~al.} 2015, \bibinfo{title}{{A Missing-link in the Supernova-GRB Connection: The Case of SN 2012ap},} \apj, 805, 187, \dodoi{10.1088/0004-637X/805/2/187}

\bibitem[{R.~A. {Chevalier}(1982){Chevalier}}]{Chevalier82self}
{Chevalier}, R.~A. 1982, \bibinfo{title}{{Self-similar solutions for the interaction of stellar ejecta with an external medium.},} \apj, 258, 790, \dodoi{10.1086/160126}

\bibitem[{R.~A. {Chevalier}(1998){Chevalier}}]{Chevalier_1998}
{Chevalier}, R.~A. 1998, \bibinfo{title}{{Synchrotron Self-Absorption in Radio Supernovae},} \apj, 499, 810, \dodoi{10.1086/305676}

\bibitem[{R.~A. {Chevalier} \& C. {Fransson}(2017){Chevalier} \& {Fransson}}]{Chevalier17}
{Chevalier}, R.~A., \& {Fransson}, C. 2017, \bibinfo{title}{{Thermal and Non-thermal Emission from Circumstellar Interaction},} in Handbook of Supernovae, ed. A.~W. {Alsabti} \& P.~{Murdin}, 875, \dodoi{10.1007/978-3-319-21846-5_34}

\bibitem[{T. {Clarke} {et~al.}(2016){Clarke}, {Kassim}, {Polisensky}, {Peters}, {Giacintucci}, \& {Hyman}}]{VLITE_paper}
{Clarke}, T., {Kassim}, N., {Polisensky}, E., {et~al.} 2016, \bibinfo{title}{{The VLA Low Band Ionospheric and Transient Experiment (VLITE): A Commensal Sky Survey},} arXiv e-prints, arXiv:1603.03080, \dodoi{10.48550/arXiv.1603.03080}

\bibitem[{A. {Corsi} \& D. {Lazzati}(2021){Corsi} \& {Lazzati}}]{Corsi_2021}
{Corsi}, A., \& {Lazzati}, D. 2021, \bibinfo{title}{{Gamma-ray burst jets in supernovae},} \nar, 92, 101614, \dodoi{10.1016/j.newar.2021.101614}

\bibitem[{A. {Corsi} {et~al.}(2014){Corsi}, {Ofek}, {Gal-Yam}, {Frail}, {Kulkarni}, {Fox}, {Kasliwal}, {Sullivan}, {Horesh}, {Carpenter}, {Maguire}, {Arcavi}, {Cenko}, {Cao}, {Mooley}, {Pan}, {Sesar}, {Sternberg}, {Xu}, {Bersier}, {James}, {Bloom}, \& {Nugent}}]{Corsi_2014}
{Corsi}, A., {Ofek}, E.~O., {Gal-Yam}, A., {et~al.} 2014, \bibinfo{title}{{A Multi-wavelength Investigation of the Radio-loud Supernova PTF11qcj and its Circumstellar Environment},} \apj, 782, 42, \dodoi{10.1088/0004-637X/782/1/42}

\bibitem[{W.~D. {Cotton}(2008){Cotton}}]{Cotton2008}
{Cotton}, W.~D. 2008, \bibinfo{title}{{Obit: A Development Environment for Astronomical Algorithms},} \pasp, 120, 439, \dodoi{10.1086/586754}

\bibitem[{F. {De Colle} {et~al.}(2022){De Colle}, {Kumar}, \& {Hoeflich}}]{De_Colle_2022}
{De Colle}, F., {Kumar}, P., \& {Hoeflich}, P. 2022, \bibinfo{title}{{The large landscape of supernova, GRB, and cocoon interactions},} \mnras, 512, 3627, \dodoi{10.1093/mnras/stac742}

\bibitem[{L. {DeMarchi} {et~al.}(2022){DeMarchi}, {Margutti}, {Dittman}, {Brunthaler}, {Milisavljevic}, {Bietenholz}, {Stauffer}, {Brethauer}, {Coppejans}, {Auchettl}, {Alexander}, {Kilpatrick}, {Bright}, {Kelley}, {Stroh}, \& {Jacobson-Gal{\'a}n}}]{DeMarchi_2022}
{DeMarchi}, L., {Margutti}, R., {Dittman}, J., {et~al.} 2022, \bibinfo{title}{{Radio Analysis of SN2004C Reveals an Unusual CSM Density Profile as a Harbinger of Core Collapse},} \apj, 938, 84, \dodoi{10.3847/1538-4357/ac8c26}

\bibitem[{D.~Z. {Dong} \& G. {Hallinan}(2023){Dong} \& {Hallinan}}]{Dong_2023}
{Dong}, D.~Z., \& {Hallinan}, G. 2023, \bibinfo{title}{{A Flat-spectrum Radio Transient at 122 Mpc Consistent with an Emerging Pulsar Wind Nebula},} \apj, 948, 119, \dodoi{10.3847/1538-4357/acc06c}

\bibitem[{R.~A.~J. {Eyles-Ferris} {et~al.}(2025){Eyles-Ferris}, {Jonker}, {Levan}, {Malesani}, {Sarin}, {Fryer}, {Rastinejad}, {Burns}, {Tanvir}, {O'Brien}, {Fong}, {Mandel}, {Gompertz}, {Kilpatrick}, {Bloemen}, {Bright}, {Carotenuto}, {Corcoran}, {Cotter}, {Groot}, {Izzo}, {Laskar}, {Martin-Carrillo}, {Palmerio}, {Ravasio}, {van Roestel}, {Saccardi}, {Starling}, {Thakur}, {Vergani}, {Vreeswijk}, {Bauer}, {Campana}, {Chac{\'o}n}, {Chrimes}, {Covino}, {van Dalen}, {D'Elia}, {De Pasquale}, {Habeeb}, {Hartmann}, {van Hoof}, {Jakobsson}, {Julakanti}, {Leloudas}, {Mata S{\'a}nchez}, {Nixon}, {Pieterse}, {Pugliese}, {Quirola-V{\'a}squez}, {Rayson}, {Salvaterra}, {Schneider}, {Torres}, \& {Zafar}}]{Eyles-Ferris_2025}
{Eyles-Ferris}, R. A.~J., {Jonker}, P.~G., {Levan}, A.~J., {et~al.} 2025, \bibinfo{title}{{The Kangaroo's First Hop: The Early Fast Cooling Phase of EP250108a/SN 2025kg},} \apjl, 988, L14, \dodoi{10.3847/2041-8213/ade1d9}

\bibitem[{D. {Foreman-Mackey} {et~al.}(2013){Foreman-Mackey}, {Hogg}, {Lang}, \& {Goodman}}]{Foreman_Mackey_2013}
{Foreman-Mackey}, D., {Hogg}, D.~W., {Lang}, D., \& {Goodman}, J. 2013, \bibinfo{title}{{emcee: The MCMC Hammer},} \pasp, 125, 306, \dodoi{10.1086/670067}

\bibitem[{C. {Fransson} {et~al.}(2005){Fransson}, {Challis}, {Chevalier}, {Filippenko}, {Kirshner}, {Kozma}, {Leonard}, {Matheson}, {Baron}, {Garnavich}, {Jha}, {Leibundgut}, {Lundqvist}, {Pun}, {Wang}, \& {Wheeler}}]{fransson05}
{Fransson}, C., {Challis}, P.~M., {Chevalier}, R.~A., {et~al.} 2005, \bibinfo{title}{{Hubble Space Telescope and Ground-based Observations of SN 1993J and SN 1998S: CNO Processing in the Progenitors},} \apj, 622, 991, \dodoi{10.1086/426495}

\bibitem[{B.~M. {Gaensler} \& P.~O. {Slane}(2006){Gaensler} \& {Slane}}]{Gaensler_2006}
{Gaensler}, B.~M., \& {Slane}, P.~O. 2006, \bibinfo{title}{{The Evolution and Structure of Pulsar Wind Nebulae},} \araa, 44, 17, \dodoi{10.1146/annurev.astro.44.051905.092528}

\bibitem[{T.~J. {Galama} {et~al.}(1998){Galama}, {Vreeswijk}, {van Paradijs}, {Kouveliotou}, {Augusteijn}, {B{\"o}hnhardt}, {Brewer}, {Doublier}, {Gonzalez}, {Leibundgut}, {Lidman}, {Hainaut}, {Patat}, {Heise}, {in't Zand}, {Hurley}, {Groot}, {Strom}, {Mazzali}, {Iwamoto}, {Nomoto}, {Umeda}, {Nakamura}, {Young}, {Suzuki}, {Shigeyama}, {Koshut}, {Kippen}, {Robinson}, {de Wildt}, {Wijers}, {Tanvir}, {Greiner}, {Pian}, {Palazzi}, {Frontera}, {Masetti}, {Nicastro}, {Feroci}, {Costa}, {Piro}, {Peterson}, {Tinney}, {Boyle}, {Cannon}, {Stathakis}, {Sadler}, {Begam}, \& {Ianna}}]{Galama_1998}
{Galama}, T.~J., {Vreeswijk}, P.~M., {van Paradijs}, J., {et~al.} 1998, \bibinfo{title}{{An unusual supernova in the error box of the {\ensuremath{\gamma}}-ray burst of 25 April 1998},} \nat, 395, 670, \dodoi{10.1038/27150}

\bibitem[{R. {Gill} \& J. {Granot}(2018){Gill} \& {Granot}}]{Gill_2018}
{Gill}, R., \& {Granot}, J. 2018, \bibinfo{title}{{Afterglow imaging and polarization of misaligned structured GRB jets and cocoons: breaking the degeneracy in GRB 170817A},} \mnras, 478, 4128, \dodoi{10.1093/mnras/sty1214}

\bibitem[{J. {Granot} \& A. {Loeb}(2003){Granot} \& {Loeb}}]{Granot_2003}
{Granot}, J., \& {Loeb}, A. 2003, \bibinfo{title}{{Radio Imaging of Gamma-Ray Burst Jets in Nearby Supernovae},} \apjl, 593, L81, \dodoi{10.1086/378262}

\bibitem[{J. {Granot} {et~al.}(2002{\natexlab{a}}){Granot}, {Panaitescu}, {Kumar}, \& {Woosley}}]{Granot_2002b}
{Granot}, J., {Panaitescu}, A., {Kumar}, P., \& {Woosley}, S.~E. 2002{\natexlab{a}}, \bibinfo{title}{{Off-Axis Afterglow Emission from Jetted Gamma-Ray Bursts},} \apjl, 570, L61, \dodoi{10.1086/340991}

\bibitem[{J. {Granot} {et~al.}(2002{\natexlab{b}}){Granot}, {Panaitescu}, {Kumar}, \& {Woosley}}]{Granot_2002_polar}
{Granot}, J., {Panaitescu}, A., {Kumar}, P., \& {Woosley}, S.~E. 2002{\natexlab{b}}, \bibinfo{title}{{Off-Axis Afterglow Emission from Jetted Gamma-Ray Bursts},} \apjl, 570, L61, \dodoi{10.1086/340991}

\bibitem[{J. {Granot} \& R. {Sari}(2002){Granot} \& {Sari}}]{Granot_2002}
{Granot}, J., \& {Sari}, R. 2002, \bibinfo{title}{{The Shape of Spectral Breaks in Gamma-Ray Burst Afterglows},} \apj, 568, 820, \dodoi{10.1086/338966}

\bibitem[{ {HI4PI Collaboration} {et~al.}(2016){HI4PI Collaboration}, {Ben Bekhti}, {Fl{\"o}er}, {Keller}, {Kerp}, {Lenz}, {Winkel}, {Bailin}, {Calabretta}, {Dedes}, {Ford}, {Gibson}, {Haud}, {Janowiecki}, {Kalberla}, {Lockman}, {McClure-Griffiths}, {Murphy}, {Nakanishi}, {Pisano}, \& {Staveley-Smith}}]{HI4PI}
{HI4PI Collaboration}, {Ben Bekhti}, N., {Fl{\"o}er}, L., {et~al.} 2016, \bibinfo{title}{{HI4PI: A full-sky H I survey based on EBHIS and GASS},} \aap, 594, A116, \dodoi{10.1051/0004-6361/201629178}

\bibitem[{A. {Jerkstrand} {et~al.}(2026){Jerkstrand}, {Milisavljevic}, \& {M{\"u}ller}}]{Jerkstrand_2026}
{Jerkstrand}, A., {Milisavljevic}, D., \& {M{\"u}ller}, B. 2026, \bibinfo{title}{{Core-collapse supernovae},} in Encyclopedia of Astrophysics, Volume 2, Vol.~2, 639--668, \dodoi{10.1016/B978-0-443-21439-4.00090-0}

\bibitem[{L. {Jewett} {et~al.}(2012){Jewett}, {Cenko}, {Li}, {Filippenko}, {Milisavljevic}, {Fesen}, {Soderberg}, {Margutti}, {Pickering}, \& {Kniazev}}]{Jewett_2012}
{Jewett}, L., {Cenko}, S.~B., {Li}, W., {et~al.} 2012, \bibinfo{title}{{Supernova 2012ap in NGC 1729 = Psn J05001372-0320512},} Central Bureau Electronic Telegrams, 3037, 1

\bibitem[{S. {Johnston} {et~al.}(2008){Johnston}, {Taylor}, {Bailes}, {Bartel}, {Baugh}, {Bietenholz}, {Blake}, {Braun}, {Brown}, {Chatterjee}, {Darling}, {Deller}, {Dodson}, {Edwards}, {Ekers}, {Ellingsen}, {Feain}, {Gaensler}, {Haverkorn}, {Hobbs}, {Hopkins}, {Jackson}, {James}, {Joncas}, {Kaspi}, {Kilborn}, {Koribalski}, {Kothes}, {Landecker}, {Lenc}, {Lovell}, {Macquart}, {Manchester}, {Matthews}, {McClure-Griffiths}, {Norris}, {Pen}, {Phillips}, {Power}, {Protheroe}, {Sadler}, {Schmidt}, {Stairs}, {Staveley-Smith}, {Stil}, {Tingay}, {Tzioumis}, {Walker}, {Wall}, \& {Wolleben}}]{ASKAP_paper}
{Johnston}, S., {Taylor}, R., {Bailes}, M., {et~al.} 2008, \bibinfo{title}{{Science with ASKAP. The Australian square-kilometre-array pathfinder},} Experimental Astronomy, 22, 151, \dodoi{10.1007/s10686-008-9124-7}

\bibitem[{A. {Kamble} {et~al.}(2016){Kamble}, {Margutti}, {Soderberg}, {Chakraborti}, {Fransson}, {Chevalier}, {Powell}, {Milisavljevic}, {Parrent}, \& {Bietenholz}}]{Kamble_2016}
{Kamble}, A., {Margutti}, R., {Soderberg}, A.~M., {et~al.} 2016, \bibinfo{title}{{Progenitors of Type IIB Supernovae in the Light of Radio and X-Rays from SN 2013DF},} \apj, 818, 111, \dodoi{10.3847/0004-637X/818/2/111}

\bibitem[{P. {Kumar} \& B. {Zhang}(2015){Kumar} \& {Zhang}}]{Kumar_2015}
{Kumar}, P., \& {Zhang}, B. 2015, \bibinfo{title}{{The physics of gamma-ray bursts \& relativistic jets},} \physrep, 561, 1, \dodoi{10.1016/j.physrep.2014.09.008}

\bibitem[{M. {Lacy} {et~al.}(2020){Lacy}, {Baum}, {Chandler}, {Chatterjee}, {Clarke}, {Deustua}, {English}, {Farnes}, {Gaensler}, {Gugliucci}, {Hallinan}, {Kent}, {Kimball}, {Law}, {Lazio}, {Marvil}, {Mao}, {Medlin}, {Mooley}, {Murphy}, {Myers}, {Osten}, {Richards}, {Rosolowsky}, {Rudnick}, {Schinzel}, {Sivakoff}, {Sjouwerman}, {Taylor}, {White}, {Wrobel}, {Andernach}, {Beasley}, {Berger}, {Bhatnager}, {Birkinshaw}, {Bower}, {Brandt}, {Brown}, {Burke-Spolaor}, {Butler}, {Comerford}, {Demorest}, {Fu}, {Giacintucci}, {Golap}, {G{\"u}th}, {Hales}, {Hiriart}, {Hodge}, {Horesh}, {Ivezi{\'c}}, {Jarvis}, {Kamble}, {Kassim}, {Liu}, {Loinard}, {Lyons}, {Masters}, {Mezcua}, {Moellenbrock}, {Mroczkowski}, {Nyland}, {O'Dea}, {O'Sullivan}, {Peters}, {Radford}, {Rao}, {Robnett}, {Salcido}, {Shen}, {Sobotka}, {Witz}, {Vaccari}, {van Weeren}, {Vargas}, {Williams}, \& {Yoon}}]{VLASS_paper}
{Lacy}, M., {Baum}, S.~A., {Chandler}, C.~J., {et~al.} 2020, \bibinfo{title}{{The Karl G. Jansky Very Large Array Sky Survey (VLASS). Science Case and Survey Design},} \pasp, 132, 035001, \dodoi{10.1088/1538-3873/ab63eb}

\bibitem[{D. {Lazzati} \& M.~C. {Begelman}(2005){Lazzati} \& {Begelman}}]{Lazzati_2005}
{Lazzati}, D., \& {Begelman}, M.~C. 2005, \bibinfo{title}{{Universal GRB Jets from Jet-Cocoon Interaction in Massive Stars},} \apj, 629, 903, \dodoi{10.1086/430877}

\bibitem[{Z. {Liu} {et~al.}(2015){Liu}, {Zhao}, {Huang}, {Wang}, {Zhang}, {Chen}, \& {Zhang}}]{Zheng_2015}
{Liu}, Z., {Zhao}, X.-L., {Huang}, F., {et~al.} 2015, \bibinfo{title}{{Optical observations of the broad-lined type Ic supernova SN 2012ap},} Research in Astronomy and Astrophysics, 15, 225, \dodoi{10.1088/1674-4527/15/2/007}

\bibitem[{A.~I. {MacFadyen} {et~al.}(2001){MacFadyen}, {Woosley}, \& {Heger}}]{MacFadyen_2001}
{MacFadyen}, A.~I., {Woosley}, S.~E., \& {Heger}, A. 2001, \bibinfo{title}{{Supernovae, Jets, and Collapsars},} \apj, 550, 410, \dodoi{10.1086/319698}

\bibitem[{R. {Margutti} \& R. {Chornock}(2021){Margutti} \& {Chornock}}]{Margutti_2021}
{Margutti}, R., \& {Chornock}, R. 2021, \bibinfo{title}{{First Multimessenger Observations of a Neutron Star Merger},} \araa, 59, 155, \dodoi{10.1146/annurev-astro-112420-030742}

\bibitem[{R. {Margutti} {et~al.}(2014){Margutti}, {Milisavljevic}, {Soderberg}, {Guidorzi}, {Morsony}, {Sanders}, {Chakraborti}, {Ray}, {Kamble}, {Drout}, {Parrent}, {Zauderer}, \& {Chomiuk}}]{Margutti_2014}
{Margutti}, R., {Milisavljevic}, D., {Soderberg}, A.~M., {et~al.} 2014, \bibinfo{title}{{Relativistic Supernovae have Shorter-lived Central Engines or More Extended Progenitors: The Case of SN 2012ap},} \apj, 797, 107, \dodoi{10.1088/0004-637X/797/2/107}

\bibitem[{R. {Margutti} {et~al.}(2017{\natexlab{a}}){Margutti}, {Kamble}, {Milisavljevic}, {Zapartas}, {de Mink}, {Drout}, {Chornock}, {Risaliti}, {Zauderer}, {Bietenholz}, {Cantiello}, {Chakraborti}, {Chomiuk}, {Fong}, {Grefenstette}, {Guidorzi}, {Kirshner}, {Parrent}, {Patnaude}, {Soderberg}, {Gehrels}, \& {Harrison}}]{Margutti_2017_14c}
{Margutti}, R., {Kamble}, A., {Milisavljevic}, D., {et~al.} 2017{\natexlab{a}}, \bibinfo{title}{{Ejection of the Massive Hydrogen-rich Envelope Timed with the Collapse of the Stripped SN 2014C},} \apj, 835, 140, \dodoi{10.3847/1538-4357/835/2/140}

\bibitem[{R. {Margutti} {et~al.}(2017{\natexlab{b}}){Margutti}, {Kamble}, {Milisavljevic}, {Zapartas}, {de Mink}, {Drout}, {Chornock}, {Risaliti}, {Zauderer}, {Bietenholz}, {Cantiello}, {Chakraborti}, {Chomiuk}, {Fong}, {Grefenstette}, {Guidorzi}, {Kirshner}, {Parrent}, {Patnaude}, {Soderberg}, {Gehrels}, \& {Harrison}}]{margutti14c}
{Margutti}, R., {Kamble}, A., {Milisavljevic}, D., {et~al.} 2017{\natexlab{b}}, \bibinfo{title}{{Ejection of the Massive Hydrogen-rich Envelope Timed with the Collapse of the Stripped SN 2014C},} \apj, 835, 140, \dodoi{10.3847/1538-4357/835/2/140}

\bibitem[{R. {Margutti} {et~al.}(2017{\natexlab{c}}){Margutti}, {Berger}, {Fong}, {Guidorzi}, {Alexander}, {Metzger}, {Blanchard}, {Cowperthwaite}, {Chornock}, {Eftekhari}, {Nicholl}, {Villar}, {Williams}, {Annis}, {Brown}, {Chen}, {Doctor}, {Frieman}, {Holz}, {Sako}, \& {Soares-Santos}}]{Margutti_2017}
{Margutti}, R., {Berger}, E., {Fong}, W., {et~al.} 2017{\natexlab{c}}, \bibinfo{title}{{The Electromagnetic Counterpart of the Binary Neutron Star Merger LIGO/Virgo GW170817. V. Rising X-Ray Emission from an Off-axis Jet},} \apjl, 848, L20, \dodoi{10.3847/2041-8213/aa9057}

\bibitem[{T. {Matsumoto} \& T. {Piran}(2023){Matsumoto} \& {Piran}}]{Matsumoto_2023}
{Matsumoto}, T., \& {Piran}, T. 2023, \bibinfo{title}{{Generalized equipartition method from an arbitrary viewing angle},} \mnras, 522, 4565, \dodoi{10.1093/mnras/stad1269}

\bibitem[{J.~C. {Mauerhan} {et~al.}(2018){Mauerhan}, {Filippenko}, {Zheng}, {Brink}, {Graham}, {Shivvers}, \& {Clubb}}]{mauerhan18}
{Mauerhan}, J.~C., {Filippenko}, A.~V., {Zheng}, W., {et~al.} 2018, \bibinfo{title}{{Stripped-envelope supernova SN 2004dk is now interacting with hydrogen-rich circumstellar material},} \mnras, 478, 5050, \dodoi{10.1093/mnras/sty1307}

\bibitem[{D. {Milisavljevic} {et~al.}(2018){Milisavljevic}, {Patnaude}, {Chevalier}, {Raymond}, {Fesen}, {Margutti}, {Conner}, \& {Banovetz}}]{dan12au}
{Milisavljevic}, D., {Patnaude}, D.~J., {Chevalier}, R.~A., {et~al.} 2018, \bibinfo{title}{{Evidence for a Pulsar Wind Nebula in the Type Ib Peculiar Supernova SN 2012au},} \apjl, 864, L36, \dodoi{10.3847/2041-8213/aadd4e}

\bibitem[{D. {Milisavljevic} {et~al.}(2015{\natexlab{a}}){Milisavljevic}, {Margutti}, {Kamble}, {Patnaude}, {Raymond}, {Eldridge}, {Fong}, {Bietenholz}, {Challis}, {Chornock}, {Drout}, {Fransson}, {Fesen}, {Grindlay}, {Kirshner}, {Lunnan}, {Mackey}, {Miller}, {Parrent}, {Sanders}, {Soderberg}, \& {Zauderer}}]{Milisavljevic_2015_14c}
{Milisavljevic}, D., {Margutti}, R., {Kamble}, A., {et~al.} 2015{\natexlab{a}}, \bibinfo{title}{{Metamorphosis of SN 2014C: Delayed Interaction between a Hydrogen Poor Core-collapse Supernova and a Nearby Circumstellar Shell},} \apj, 815, 120, \dodoi{10.1088/0004-637X/815/2/120}

\bibitem[{D. {Milisavljevic} {et~al.}(2015{\natexlab{b}}){Milisavljevic}, {Margutti}, {Parrent}, {Soderberg}, {Fesen}, {Mazzali}, {Maeda}, {Sanders}, {Cenko}, {Silverman}, {Filippenko}, {Kamble}, {Chakraborti}, {Drout}, {Kirshner}, {Pickering}, {Kawabata}, {Hattori}, {Hsiao}, {Stritzinger}, {Marion}, {Vinko}, \& {Wheeler}}]{Milisavljevic_2015}
{Milisavljevic}, D., {Margutti}, R., {Parrent}, J.~T., {et~al.} 2015{\natexlab{b}}, \bibinfo{title}{{The Broad-lined Type Ic SN 2012ap and the Nature of Relativistic Supernovae Lacking a Gamma-Ray Burst Detection},} \apj, 799, 51, \dodoi{10.1088/0004-637X/799/1/51}

\bibitem[{D. {Milisavljevic} {et~al.}(2015{\natexlab{c}}){Milisavljevic}, {Margutti}, {Kamble}, {Patnaude}, {Raymond}, {Eldridge}, {Fong}, {Bietenholz}, {Challis}, {Chornock}, {Drout}, {Fransson}, {Fesen}, {Grindlay}, {Kirshner}, {Lunnan}, {Mackey}, {Miller}, {Parrent}, {Sanders}, {Soderberg}, \& {Zauderer}}]{milisav14c}
{Milisavljevic}, D., {Margutti}, R., {Kamble}, A., {et~al.} 2015{\natexlab{c}}, \bibinfo{title}{{Metamorphosis of SN 2014C: Delayed Interaction between a Hydrogen Poor Core-collapse Supernova and a Nearby Circumstellar Shell},} \apj, 815, 120, \dodoi{10.1088/0004-637X/815/2/120}

\bibitem[{K. {Murase} {et~al.}(2021){Murase}, {Omand}, {Coppejans}, {Nagai}, {Bower}, {Chornock}, {Fox}, {Kashiyama}, {Law}, {Margutti}, \& {M{\'e}sz{\'a}ros}}]{Murase_2021}
{Murase}, K., {Omand}, C. M.~B., {Coppejans}, D.~L., {et~al.} 2021, \bibinfo{title}{{ALMA and NOEMA constraints on synchrotron nebular emission from embryonic superluminous supernova remnants and radio-gamma-ray connection},} \mnras, 508, 44, \dodoi{10.1093/mnras/stab2506}

\bibitem[{T. {Murphy} {et~al.}(2013){Murphy}, {Chatterjee}, {Kaplan}, {Banyer}, {Bell}, {Bignall}, {Bower}, {Cameron}, {Coward}, {Cordes}, {Croft}, {Curran}, {Djorgovski}, {Farrell}, {Frail}, {Gaensler}, {Galloway}, {Gendre}, {Green}, {Hancock}, {Johnston}, {Kamble}, {Law}, {Lazio}, {Lo}, {Macquart}, {Rea}, {Rebbapragada}, {Reynolds}, {Ryder}, {Schmidt}, {Soria}, {Stairs}, {Tingay}, {Torkelsson}, {Wagstaff}, {Walker}, {Wayth}, \& {Williams}}]{VAST_paper}
{Murphy}, T., {Chatterjee}, S., {Kaplan}, D.~L., {et~al.} 2013, \bibinfo{title}{{VAST: An ASKAP Survey for Variables and Slow Transients},} \pasa, 30, e006, \dodoi{10.1017/pasa.2012.006}

\bibitem[{E. {Nakar} \& T. {Piran}(2017){Nakar} \& {Piran}}]{Nakar_2017}
{Nakar}, E., \& {Piran}, T. 2017, \bibinfo{title}{{The Observable Signatures of GRB Cocoons},} \apj, 834, 28, \dodoi{10.3847/1538-4357/834/1/28}

\bibitem[{J.~B. {Oke} {et~al.}(1995){Oke}, {Cohen}, {Carr}, {Cromer}, {Dingizian}, {Harris}, {Labrecque}, {Lucinio}, {Schaal}, {Epps}, \& {Miller}}]{lrisref}
{Oke}, J.~B., {Cohen}, J.~G., {Carr}, M., {et~al.} 1995, \bibinfo{title}{{The Keck Low-Resolution Imaging Spectrometer},} \pasp, 107, 375, \dodoi{10.1086/133562}

\bibitem[{C.~M.~B. {Omand} \& A. {Jerkstrand}(2023){Omand} \& {Jerkstrand}}]{Omand_2023}
{Omand}, C.~M.~B., \& {Jerkstrand}, A. 2023, \bibinfo{title}{{Toward nebular spectral modeling of magnetar-powered supernovae},} \aap, 673, A107, \dodoi{10.1051/0004-6361/202245406}

\bibitem[{D.~E. {Osterbrock} \& G.~J. {Ferland}(2006){Osterbrock} \& {Ferland}}]{agn2}
{Osterbrock}, D.~E., \& {Ferland}, G.~J. 2006, {Astrophysics of gaseous nebulae and active galactic nuclei}

\bibitem[{A.~G. {Pacholczyk}(1970){Pacholczyk}}]{Pacholczyk_1970}
{Pacholczyk}, A.~G. 1970, {Radio astrophysics. Nonthermal processes in galactic and extragalactic sources} (Series of Books in Astronomy and Astrophysics, San Francisco: Freeman, 1970)

\bibitem[{N.~T. {Palliyaguru} {et~al.}(2021){Palliyaguru}, {Corsi}, {P{\'e}rez-Torres}, {Varenius}, \& {Van Eerten}}]{Palliyaguru_2021}
{Palliyaguru}, N.~T., {Corsi}, A., {P{\'e}rez-Torres}, M., {Varenius}, E., \& {Van Eerten}, H. 2021, \bibinfo{title}{{VLBI Observations of Supernova PTF11qcj: Direct Constraints on the Size of the Radio Ejecta},} \apj, 910, 16, \dodoi{10.3847/1538-4357/abe1c9}

\bibitem[{N.~T. {Palliyaguru} {et~al.}(2019){Palliyaguru}, {Corsi}, {Frail}, {Vink{\'o}}, {Wheeler}, {Gal-Yam}, {Cenko}, {Kulkarni}, \& {Kasliwal}}]{Palliyaguru_2019}
{Palliyaguru}, N.~T., {Corsi}, A., {Frail}, D.~A., {et~al.} 2019, \bibinfo{title}{{The Double-peaked Radio Light Curve of Supernova PTF11qcj},} \apj, 872, 201, \dodoi{10.3847/1538-4357/aaf64d}

\bibitem[{G. {Pignata} {et~al.}(2011){Pignata}, {Stritzinger}, {Soderberg}, {Mazzali}, {Phillips}, {Morrell}, {Anderson}, {Boldt}, {Campillay}, {Contreras}, {Folatelli}, {F{\"o}rster}, {Gonz{\'a}lez}, {Hamuy}, {Krzeminski}, {Maza}, {Roth}, {Salgado}, {Levesque}, {Rest}, {Crain}, {Foster}, {Haislip}, {Ivarsen}, {LaCluyze}, {Nysewander}, \& {Reichart}}]{Pignata_2011}
{Pignata}, G., {Stritzinger}, M., {Soderberg}, A., {et~al.} 2011, \bibinfo{title}{{SN 2009bb: A Peculiar Broad-lined Type Ic Supernova},} \apj, 728, 14, \dodoi{10.1088/0004-637X/728/1/14}

\bibitem[{T. {Piran}(2004){Piran}}]{Piran_2004}
{Piran}, T. 2004, \bibinfo{title}{{The physics of gamma-ray bursts},} Reviews of Modern Physics, 76, 1143, \dodoi{10.1103/RevModPhys.76.1143}

\bibitem[{E. {Polisensky} {et~al.}(2016){Polisensky}, {Lane}, {Hyman}, {Kassim}, {Giacintucci}, {Clarke}, {Cotton}, {Cleland}, \& {Frail}}]{Polisensky2016}
{Polisensky}, E., {Lane}, W.~M., {Hyman}, S.~D., {et~al.} 2016, \bibinfo{title}{{Exploring the Transient Radio Sky with VLITE: Early Results},} \apj, 832, 60, \dodoi{10.3847/0004-637X/832/1/60}

\bibitem[{S.~M. {Rahaman} {et~al.}(2026){Rahaman}, {Granot}, \& {Beniamini}}]{Minhajur_2026}
{Rahaman}, S.~M., {Granot}, J., \& {Beniamini}, P. 2026, \bibinfo{title}{{The Deep Newtonian Regime in Late-Time Blast Waves: Inevitable Transition and Distinct Flux Signatures},} arXiv e-prints, arXiv:2604.23567, \dodoi{10.48550/arXiv.2604.23567}

\bibitem[{S.~P. {Reynolds} \& R.~A. {Chevalier}(1984){Reynolds} \& {Chevalier}}]{Reynolds_1984}
{Reynolds}, S.~P., \& {Chevalier}, R.~A. 1984, \bibinfo{title}{{Evolution of pulsar-driven supernova remnants.},} \apj, 278, 630, \dodoi{10.1086/161831}

\bibitem[{K. {Rose} {et~al.}(2024){Rose}, {Horesh}, {Murphy}, {Kaplan}, {Sfaradi}, {Ryder}, {Aloisi}, {Dobie}, {Driessen}, {Fender}, {Green}, {Leung}, {Lenc}, {Qiu}, \& {Williams-Baldwin}}]{Rose_2024}
{Rose}, K., {Horesh}, A., {Murphy}, T., {et~al.} 2024, \bibinfo{title}{{Late-time supernovae radio re-brightening in the VAST pilot survey},} \mnras, 534, 3853, \dodoi{10.1093/mnras/stae2289}

\bibitem[{R. {Sari} {et~al.}(1998){Sari}, {Piran}, \& {Narayan}}]{Sari_1998}
{Sari}, R., {Piran}, T., \& {Narayan}, R. 1998, \bibinfo{title}{{Spectra and Light Curves of Gamma-Ray Burst Afterglows},} \apjl, 497, L17, \dodoi{10.1086/311269}

\bibitem[{G. {Schroeder} {et~al.}(2025){Schroeder}, {Ho}, {Dastidar}, {Modjaz}, {Corsi}, \& {Duffell}}]{Scheroder_2025}
{Schroeder}, G., {Ho}, A. Y.~Q., {Dastidar}, R.~G., {et~al.} 2025, \bibinfo{title}{{A Late-time Radio Search for Highly Off-axis Jets from PTF Broad-lined Ic Supernovae in GRB-like Host Galaxy Environments},} arXiv e-prints, arXiv:2507.15928, \dodoi{10.48550/arXiv.2507.15928}

\bibitem[{I. {Sfaradi} {et~al.}(2025{\natexlab{a}}){Sfaradi}, {Horesh}, {Fender}, {Rhodes}, {Bright}, {Williams-Baldwin}, \& {Green}}]{Sfaradi_2025}
{Sfaradi}, I., {Horesh}, A., {Fender}, R., {et~al.} 2025{\natexlab{a}}, \bibinfo{title}{{The Observed Phase Space of Mass-loss History from Massive Stars Based on Radio Observations of a Large Supernova Sample},} \apj, 979, 189, \dodoi{10.3847/1538-4357/ad9e93}

\bibitem[{I. {Sfaradi} {et~al.}(2024){Sfaradi}, {Beniamini}, {Horesh}, {Piran}, {Bright}, {Rhodes}, {Williams}, {Fender}, {Leung}, {Murphy}, \& {Green}}]{Sfaradi_2024}
{Sfaradi}, I., {Beniamini}, P., {Horesh}, A., {et~al.} 2024, \bibinfo{title}{{An off-axis relativistic jet seen in the long lasting delayed radio flare of the TDE AT 2018hyz},} \mnras, 527, 7672, \dodoi{10.1093/mnras/stad3717}

\bibitem[{I. {Sfaradi} {et~al.}(2025{\natexlab{b}}){Sfaradi}, {Margutti}, {Chornock}, {Alexander}, {Metzger}, {Beniamini}, {Duran}, {Yao}, {Horesh}, {Farah}, {Berger}, {A.~J.}, {Cendes}, {Eftekhari}, {Fender}, {Franz}, {Green}, {Hammerstein}, {Lu}, {Wiston}, {Bernstein}, {Bright}, {Christy}, {Cruz}, {DeBoer}, {Golay}, {Goodwin}, {Gurwell}, {Keating}, {Laskar}, {Miller-Jones}, {Pollak}, {Rao}, {Siemion}, {Sheikh}, {Shoval}, \& {van Velzen}}]{Sfaradi_2025_tvd}
{Sfaradi}, I., {Margutti}, R., {Chornock}, R., {et~al.} 2025{\natexlab{b}}, \bibinfo{title}{{The First Radio-bright Off-nuclear Tidal Disruption Event AT 2024tvd Reveals the Fastest-evolving Double-peaked Radio Emission},} \apjl, 992, L18, \dodoi{10.3847/2041-8213/ae0a26}

\bibitem[{J.~M. {Silverman} {et~al.}(2012){Silverman}, {Foley}, {Filippenko}, {Ganeshalingam}, {Barth}, {Chornock}, {Griffith}, {Kong}, {Lee}, {Leonard}, {Matheson}, {Miller}, {Steele}, {Barris}, {Bloom}, {Cobb}, {Coil}, {Desroches}, {Gates}, {Ho}, {Jha}, {Kandrashoff}, {Li}, {Mandel}, {Modjaz}, {Moore}, {Mostardi}, {Papenkova}, {Park}, {Perley}, {Poznanski}, {Reuter}, {Scala}, {Serduke}, {Shields}, {Swift}, {Tonry}, {Van Dyk}, {Wang}, \& {Wong}}]{silverman2012}
{Silverman}, J.~M., {Foley}, R.~J., {Filippenko}, A.~V., {et~al.} 2012, \bibinfo{title}{{Berkeley Supernova Ia Program - I. Observations, data reduction and spectroscopic sample of 582 low-redshift Type Ia supernovae},} \mnras, 425, 1789, \dodoi{10.1111/j.1365-2966.2012.21270.x}

\bibitem[{S.~J. {Smartt}(2009){Smartt}}]{Smartt_2009}
{Smartt}, S.~J. 2009, \bibinfo{title}{{Progenitors of Core-Collapse Supernovae},} \araa, 47, 63, \dodoi{10.1146/annurev-astro-082708-101737}

\bibitem[{N. {Smith} {et~al.}(2009){Smith}, {Silverman}, {Chornock}, {Filippenko}, {Wang}, {Li}, {Ganeshalingam}, {Foley}, {Rex}, \& {Steele}}]{smith05ip}
{Smith}, N., {Silverman}, J.~M., {Chornock}, R., {et~al.} 2009, \bibinfo{title}{{Coronal Lines and Dust Formation in SN 2005ip: Not the Brightest, but the Hottest Type IIn Supernova},} \apj, 695, 1334, \dodoi{10.1088/0004-637X/695/2/1334}

\bibitem[{A.~M. {Soderberg} {et~al.}(2006){Soderberg}, {Nakar}, {Berger}, \& {Kulkarni}}]{Soderberg_2006}
{Soderberg}, A.~M., {Nakar}, E., {Berger}, E., \& {Kulkarni}, S.~R. 2006, \bibinfo{title}{{Late-Time Radio Observations of 68 Type Ibc Supernovae: Strong Constraints on Off-Axis Gamma-Ray Bursts},} \apj, 638, 930, \dodoi{10.1086/499121}

\bibitem[{A.~M. {Soderberg} {et~al.}(2010){Soderberg}, {Chakraborti}, {Pignata}, {Chevalier}, {Chandra}, {Ray}, {Wieringa}, {Copete}, {Chaplin}, {Connaughton}, {Barthelmy}, {Bietenholz}, {Chugai}, {Stritzinger}, {Hamuy}, {Fransson}, {Fox}, {Levesque}, {Grindlay}, {Challis}, {Foley}, {Kirshner}, {Milne}, \& {Torres}}]{Soderberg_2010}
{Soderberg}, A.~M., {Chakraborti}, S., {Pignata}, G., {et~al.} 2010, \bibinfo{title}{{A relativistic type Ibc supernova without a detected {\ensuremath{\gamma}}-ray burst},} \nat, 463, 513, \dodoi{10.1038/nature08714}

\bibitem[{M. {Solar} {et~al.}(2024){Solar}, {Micha{\l}owski}, {Nadolny}, {Galbany}, {Hjorth}, {Zapartas}, {Sollerman}, {Hunt}, {Klose}, {Koprowski}, {Le{\'s}niewska}, {Ma{\l}kowski}, {Nicuesa Guelbenzu}, {Ryzhov}, {Savaglio}, {Schady}, {Schulze}, {de Ugarte Postigo}, {Vergani}, {Watson}, \& {Wr{\'o}blewski}}]{Martin_2024}
{Solar}, M., {Micha{\l}owski}, M.~J., {Nadolny}, J., {et~al.} 2024, \bibinfo{title}{{Binary progenitor systems for Type Ic supernovae},} Nature Communications, 15, 7667, \dodoi{10.1038/s41467-024-51863-z}

\bibitem[{G.~P. {Srinivasaragavan} {et~al.}(2025){Srinivasaragavan}, {Hamidani}, {Schroeder}, {Sarin}, {Ho}, {Piro}, {Cenko}, {Anand}, {Sollerman}, {Perley}, {Maeda}, {O'Connor}, {Kuncarayakti}, {Miller}, {Ahumada}, {Vail}, {Duffell}, {Dastidar}, {Andreoni}, {Bochenek}, {Brennan}, {Carney}, {Chen}, {Freeburn}, {Gal-Yam}, {Jacobson-Gal{\'a}n}, {Kasliwal}, {Li}, {Li}, {Sravan}, \& {Warshofsky}}]{Srinivasaragavan_2025}
{Srinivasaragavan}, G.~P., {Hamidani}, H., {Schroeder}, G., {et~al.} 2025, \bibinfo{title}{{EP250108a/SN 2025kg: A Jet-driven Stellar Explosion Interacting with Circumstellar Material},} \apjl, 988, L60, \dodoi{10.3847/2041-8213/ade870}

\bibitem[{S. {Srivastav} {et~al.}(2025){Srivastav}, {Chen}, {Gillanders}, {Rhodes}, {Smartt}, {Huber}, {Aryan}, {Yang}, {Beri}, {Cooper}, {Nicholl}, {Smith}, {Stevance}, {Carotenuto}, {Chambers}, {Aamer}, {Angus}, {Fulton}, {Moore}, {Smith}, {Young}, {de Boer}, {Gao}, {Lin}, {Lowe}, {Magnier}, {Minguez}, {Pan}, \& {Wainscoat}}]{Srivastav_2025}
{Srivastav}, S., {Chen}, T.-W., {Gillanders}, J.~H., {et~al.} 2025, \bibinfo{title}{{Identification of the Optical Counterpart of the Fast X-Ray Transient EP240414a},} \apjl, 978, L21, \dodoi{10.3847/2041-8213/ad9c75}

\bibitem[{K.~Z. {Stanek} {et~al.}(2003){Stanek}, {Matheson}, {Garnavich}, {Martini}, {Berlind}, {Caldwell}, {Challis}, {Brown}, {Schild}, {Krisciunas}, {Calkins}, {Lee}, {Hathi}, {Jansen}, {Windhorst}, {Echevarria}, {Eisenstein}, {Pindor}, {Olszewski}, {Harding}, {Holland}, \& {Bersier}}]{Stanek_2003}
{Stanek}, K.~Z., {Matheson}, T., {Garnavich}, P.~M., {et~al.} 2003, \bibinfo{title}{{Spectroscopic Discovery of the Supernova 2003dh Associated with GRB 030329},} \apjl, 591, L17, \dodoi{10.1086/376976}

\bibitem[{M.~C. {Stroh} {et~al.}(2021){Stroh}, {Terreran}, {Coppejans}, {Bright}, {Margutti}, {Bietenholz}, {De Colle}, {DeMarchi}, {Duran}, {Milisavljevic}, {Murase}, {Paterson}, \& {Williams}}]{Stroh_2021}
{Stroh}, M.~C., {Terreran}, G., {Coppejans}, D.~L., {et~al.} 2021, \bibinfo{title}{{Luminous Late-time Radio Emission from Supernovae Detected by the Karl G. Jansky Very Large Array Sky Survey (VLASS)},} \apjl, 923, L24, \dodoi{10.3847/2041-8213/ac375e}

\bibitem[{B.~P. {Thomas} {et~al.}(2022){Thomas}, {Wheeler}, {Dwarkadas}, {Stockdale}, {Vink{\'o}}, {Pooley}, {Xu}, {Zeimann}, \& {MacQueen}}]{thomas22}
{Thomas}, B.~P., {Wheeler}, J.~C., {Dwarkadas}, V.~V., {et~al.} 2022, \bibinfo{title}{{Seven Years of SN 2014C: A Multiwavelength Synthesis of an Extraordinary Supernova},} \apj, 930, 57, \dodoi{10.3847/1538-4357/ac5fa6}

\bibitem[{S. {Tinyanont} {et~al.}(2025){Tinyanont}, {Fox}, {Shahbandeh}, {Temim}, {Williams}, {Wangnok}, {Rest}, {Lau}, {Maeda}, {Jencson}, {Auchettl}, {Filippenko}, {Larison}, {Ashall}, {Brink}, {Davis}, {Dessart}, {Foley}, {Galbany}, {Grayling}, {Johansson}, {Kasliwal}, {Lane}, {LeBaron}, {Milisavljevic}, {Rho}, {Sakon}, {Sarangi}, {Szalai}, {Taggart}, {Van Dyk}, {Wang}, {Yang}, {Zheng}, \& {Zs{\'\i}ros}}]{tinyanont25}
{Tinyanont}, S., {Fox}, O.~D., {Shahbandeh}, M., {et~al.} 2025, \bibinfo{title}{{Large Cold Dust Reservoir Revealed in Transitional SN Ib 2014C by James Webb Space Telescope Mid-infrared Spectroscopy},} \apj, 985, 198, \dodoi{10.3847/1538-4357/adccc0}

\bibitem[{M. {Turatto} {et~al.}(1993){Turatto}, {Cappellaro}, {Danziger}, {Benetti}, {Gouiffes}, \& {della Valle}}]{turatto93}
{Turatto}, M., {Cappellaro}, E., {Danziger}, I.~J., {et~al.} 1993, \bibinfo{title}{{The type II supernova 1988Z in MCG +03-28-022 : increasingevidence of interaction of supernova ejecta with a circumstellar wind.},} \mnras, 262, 128, \dodoi{10.1093/mnras/262.1.128}

\bibitem[{Y. {Wang} {et~al.}(2026){Wang}, {Chen}, \& {Zhang}}]{Wang_2026}
{Wang}, Y., {Chen}, C., \& {Zhang}, B. 2026, \bibinfo{title}{{VegasAfterglow: A high-performance framework for gamma-ray burst afterglows},} Journal of High Energy Astrophysics, 50, 100490, \dodoi{10.1016/j.jheap.2025.100490}

\bibitem[{K.~W. {Weiler} {et~al.}(2002){Weiler}, {Panagia}, {Montes}, \& {Sramek}}]{Weiler_2002}
{Weiler}, K.~W., {Panagia}, N., {Montes}, M.~J., \& {Sramek}, R.~A. 2002, \bibinfo{title}{{Radio Emission from Supernovae and Gamma-Ray Bursters},} \araa, 40, 387, \dodoi{10.1146/annurev.astro.40.060401.093744}

\bibitem[{K.~W. {Weiler} {et~al.}(1986){Weiler}, {Sramek}, {Panagia}, {van der Hulst}, \& {Salvati}}]{Weiler_1986}
{Weiler}, K.~W., {Sramek}, R.~A., {Panagia}, N., {van der Hulst}, J.~M., \& {Salvati}, M. 1986, \bibinfo{title}{{Radio Supernovae},} \apj, 301, 790, \dodoi{10.1086/163944}

\bibitem[{S.~E. {Woosley} \& J.~S. {Bloom}(2006){Woosley} \& {Bloom}}]{Woosley_2006}
{Woosley}, S.~E., \& {Bloom}, J.~S. 2006, \bibinfo{title}{{The Supernova Gamma-Ray Burst Connection},} \araa, 44, 507, \dodoi{10.1146/annurev.astro.43.072103.150558}

\bibitem[{W. {Yuan} {et~al.}(2015){Yuan}, {Zhang}, {Feng}, {Zhang}, {Ling}, {Zhao}, {Deng}, {Qiu}, {Osborne}, {O'Brien}, {Willingale}, {Lapington}, {Fraser}, \& {the Einstein Probe team}}]{Yuan_2015}
{Yuan}, W., {Zhang}, C., {Feng}, H., {et~al.} 2015, \bibinfo{title}{{Einstein Probe - a small mission to monitor and explore the dynamic X-ray Universe},} arXiv e-prints, arXiv:1506.07735, \dodoi{10.48550/arXiv.1506.07735}

\bibitem[{J.-H. {Zheng} {et~al.}(2025){Zheng}, {Zhu}, {Lu}, \& {Zhang}}]{Zheng_2025}
{Zheng}, J.-H., {Zhu}, J.-P., {Lu}, W., \& {Zhang}, B. 2025, \bibinfo{title}{{EP240414a: Off-axis View of a Jet-cocoon System from an Expanded Progenitor Star},} \apj, 985, 21, \dodoi{10.3847/1538-4357/adc993}

\end{thebibliography}
\bibliographystyle{aasjournalv7}

\end{document}